\documentclass[journal]{IEEEtran}
\usepackage{amsmath,amssymb,amsfonts,amsthm}
\usepackage{enumerate}
\usepackage{pifont}
\usepackage{algorithmic}
\usepackage{tikz}
\usepackage{cases}
\usepackage{color}
\usepackage{bm}
\usepackage{multirow}
\usepackage{fix-cm}
\usepackage{textcomp}
\usepackage{epstopdf}
\usepackage{stmaryrd}
\usepackage{makecell}
\usepackage{chngpage}
\usepackage{subfigure}
\usepackage{color}
\usepackage{gensymb}
\usepackage{graphicx,graphics}
\usepackage{algorithm}
\usepackage{cite}
\usepackage{booktabs}
\usepackage[hidelinks]{hyperref}

\theoremstyle{plain}
\makeatletter
\g@addto@macro\th@plain{\thm@headpunct{}}
\makeatother

 \def\BibTeX{{\rm
B\kern-.05em{\sc i\kern-.025em b}\kern-.08em
T\kern-.1667em\lower.7ex\hbox{E}\kern-.125emX}}
\begin{document}
\title{Joint Synchronization and Sensing in Networked ISAC via Structured Canonical Polyadic Decomposition}
\author{Lin Chen, Yifan Liang, and Hongbin Li,~\IEEEmembership{Fellow,~IEEE}
\thanks{This work was supported in part by the National Science Foundation under
Grants CNS-2533158, CCF-2316865, and ECCS-2332534.
\emph{(Corresponding author: Hongbin Li.)}} \thanks{The authors are with the
Department of Electrical and Computer Engineering, Stevens Institute of
Technology, Hoboken, NJ 07030, USA (e-mail: lchen53@stevens.edu;
yliang33@stevens.edu; hli@stevens.edu).}}

\markboth{}
{Shell \MakeLowercase{\textit{et al.}}: Bare Demo of IEEEtran.cls for IEEE
Journals}
\maketitle

\begin{abstract}
Networked integrated sensing and communication (ISAC) offers significant potential for next-generation wireless systems. By exploiting spatial diversity through the cooperation of multiple base stations (BSs), this architecture expands coverage and achieves enhanced sensing performance. However, accurate sensing in networked ISAC requires time-frequency synchronization among BSs. Existing synchronization methods for networked ISAC suffer from inter-path interference caused by sensing channel compression. To address this problem, this paper proposes a structured canonical polyadic decomposition (SCPD) algorithm that effectively separates the multipath components of the sensing channel. Benefiting from this separation, SCPD achieves joint network-level synchronization and multi-target parameter estimation. We establish theoretical identifiability conditions for SCPD and show that it asymptotically achieves the Cram\'{e}r-Rao bound. Furthermore, by incorporating parameters estimated from different BS pairs, we propose a multi-target tracking algorithm designed for the continuous operation of the system. The proposed algorithm tracks both the trajectories and velocities of moving targets by leveraging geometric diversity. Utilizing tracking results from the previous snapshot, an adaptive beamforming scheme is also developed to improve tracking performance in the next snapshot. Simulation results demonstrate that the proposed algorithms achieve superior accuracy and outlier robustness for both synchronization and sensing in networked ISAC, outperforming traditional approaches.
\end{abstract}
\begin{IEEEkeywords}
Networked integrated sensing and communication (ISAC), synchronization, target sensing, adaptive beamforming, canonical polyadic decomposition (CPD).
\end{IEEEkeywords}

\IEEEpeerreviewmaketitle

\section{Introduction}
\IEEEPARstart{N}{etworked} integrated sensing and communication (ISAC) has emerged as a promising architecture for sixth-generation (6G) wireless systems\cite{overview2,overview,5GNR,nist2025cooperative,wei2023integrated,wang2025cooperative,xia2026joint,beam1,est1,offset}. In a networked ISAC configuration, multiple geographically distributed base stations (BSs) cooperatively provide communication services, while simultaneously reusing the same spectrum, hardware, and signaling resources to sense the environment. This cooperative framework enables the localization of both user equipments (UEs) and non-cooperative targets. Compared with single-cell ISAC, networked ISAC observes targets from diverse viewpoints, thus providing improved spatial diversity and extending sensing coverage. Moreover, since BSs are already widely deployed in cellular networks, networked ISAC offers a practical paradigm for large-scale environmental perception without requiring a dedicated sensing infrastructure. These capabilities make networked ISAC attractive for broad applications such as autonomous driving, industrial automation, and unmanned aerial vehicle monitoring \cite{overview2,overview,5GNR,nist2025cooperative,wang2025cooperative,wei2023integrated}.
\par Realizing these capabilities in networked ISAC requires the cooperating BSs to operate on a common time-frequency reference. In practice, each BS is driven by an independent local oscillator, which induces timing offsets (TOs) and carrier frequency offsets (CFOs) among the BSs\cite{traditional,cox2005time,level,sensingModel,est1,offset}. If left uncompensated, these offsets distort the bistatic delay and Doppler measurements, degrading both communication cooperation and sensing accuracy. Hence, accurate time-frequency synchronization is a fundamental prerequisite for reliable networked ISAC. Traditional synchronization techniques based on the master-slave architecture \cite{traditional,cox2005time} provide microsecond-level precision in terms of TO estimation, while global positioning system (GPS)-aided methods \cite{sensingModel,level} offer about 100 ns precision. However, such synchronization precision remains insufficient for sub-meter target localization. As an enhancement, the time synchronization approach proposed in \cite{est1} achieves an accuracy of about 1 ns. Nevertheless, this approach requires the prior knowledge of the locations of surrounding targets. Any deviation in target localization inevitably degrades the TO estimation performance.
\par Aiming to decouple offset estimation from target localization, a super-resolution offset estimation method by matrix pencil (SOE-MP) was proposed in \cite{offset}. SOE-MP compresses the time-frequency sensing channel into a vector via the discrete Fourier transform (DFT). Then, it exploits an offset reciprocity in bistatic links \cite{offset,radar} to estimate TOs and CFOs. However, the DFT-based compression destroys the multi-dimensional structure of the sensing channel, causing severe information loss. Moreover, this channel compression introduces inter-path interference, resulting in unreliable offset estimation. Specifically, the compressed vector is a superposition of multiple paths, making the desired path carrying TO or CFO information indistinguishable from others.
\par To avoid information loss and inter-path interference, canonical polyadic decomposition (CPD)\cite{zhou,RZ,chen2026tensor,syncletter,Vandermonde,qian2018tensor} can be used to model multi-dimensional signals in ISAC systems. Specifically, sensing signals received across different arrays, subcarriers, and symbols can be stacked to form a tensor, preserving the multipath parameter information of the sensing channel. Notably, such a tensor admits a unique CPD under mild conditions\cite{zhou,RZ,chen2026tensor,syncletter,Vandermonde,qian2018tensor}, enabling the isolation of multipath components of the channel into distinct factor matrices. However, traditional CPD\cite{zhou} neglects the prior structure of these factor matrices. According to the sensing signal model of ISAC\cite{RZ,chen2026tensor,syncletter}, factor matrices in CPD exhibit a Vandermonde structure\cite{Vandermonde,qian2018tensor}. This structure is exploited in the CPD with Vandermonde factor matrix (CP-VDM) method\cite{Vandermonde} to enhance standard CPD. CP-VDM decomposes the tensor observation into orthogonal signal and noise subspaces, and then estimates the Vandermonde factor matrices within the signal subspace. It achieves unbiased estimation in the noiseless scenario, but suffers from performance degeneration in low signal-to-noise ratio (SNR) regimes, where accurate signal subspace estimation is challenging\cite{qian2018tensor,chen2026tensor}.
\par By isolating multipath components in the sensing channel, the tensor-based framework employing CPD can jointly estimate not only offset parameters but also target parameters, including angle, bistatic delay, and Doppler shift. Since tensor-based estimators effectively preserve the multi-dimensional structure of the signals, they achieve enhanced estimation accuracy in ISAC systems compared to matrix-based estimators, such as two-dimensional DFT (2D-DFT)\cite{hu2024joint}, multiple signal classification (MUSIC)\cite{musicDOA}, and estimation of signal parameters via rotational variation technique (ESPRIT)\cite{xiang2023esprit}. Among these matrix-based approaches, 2D-DFT is computationally efficient but suffers from limited resolution. Super-resolution algorithms like MUSIC and ESPRIT offer higher resolution than 2D-DFT, yet they are ill-suited for characterizing the multi-dimensional signal structure\cite{chen2026tensor}. Additionally, compressed sensing-based methods, such as orthogonal matching pursuit (OMP)\cite{gao2022integrated,duong2025sparse}, have been applied to ISAC systems, but they are computationally expensive for joint parameter estimation. Specifically, OMP becomes infeasible in large-scale scenarios due to the need to construct a multi-dimensional dictionary, whose computational complexity scales exponentially with parameter resolution\cite{chen2026tensor,RZ,musicDOA}.
\par Importantly, the aforementioned target sensing works\cite{musicDOA,hu2024joint,gao2022integrated,duong2025sparse,RZ,chen2026tensor,xiang2023esprit} focus on single-cell ISAC systems, leaving target sensing in networked ISAC largely unexplored. In a networked ISAC system, each BS pair can estimate target angle and delay parameters, enabling independent target localization. Nevertheless, localization accuracy can be significantly enhanced via inter-BS collaboration, which exploits distributed viewpoints and improves spatial diversity. Furthermore, a single BS pair cannot recover the 2D velocity of a target, as its estimated Doppler shift yields only the radial velocity component but lacks transverse velocity information. Fortunately, a networked ISAC system observes the same target from multiple spatially separated BS pairs. Such multi-baseline measurements provide geometric diversity and enable 2D (or even higher-dimensional) velocity recovery\cite{nist2025cooperative,xia2026joint}. An analogous framework can be found in distributed multiple-input multiple-output (MIMO) radar systems\cite{wang2011moving,sadeghi2021target,xiong2022distributed,lai2023joint}, which achieve geometric gains for target localization but require dedicated waveforms and assume perfect synchronization.
\par To overcome the limitations of existing synchronization methods\cite{traditional,cox2005time,level,sensingModel,est1,offset} and target sensing methods\cite{musicDOA,hu2024joint,gao2022integrated,duong2025sparse,RZ,chen2026tensor,xiang2023esprit}, this paper develops an integrated synchronization and sensing framework for networked ISAC. The main contributions of this work are summarized as follows.
\begin{itemize}
\item We propose a tensor-based framework for joint time-frequency synchronization and target sensing in networked ISAC. In this framework, each BS pair independently estimates offset and target parameters from the received sensing signals. The proposed estimator isolates multipath components in the sensing channel to avoid inter-path interference. Furthermore, by fusing estimates from different BS pairs, the proposed framework tracks the trajectories and velocities of moving targets over multiple snapshots. Additionally, an adaptive beamforming scheme is developed to ensure tracking performance.
\item We propose a structured CPD (SCPD) algorithm for offset and target parameter estimation. SCPD enforces Vandermonde structural constraints on the factor matrices to improve decomposition accuracy. The resulting optimization problem is solved in a constrained alternating least-squares (CALS) framework, demonstrating enhanced noise robustness over traditional subspace-based solutions like CP-VDM. As a maximum likelihood estimator under mild assumptions, SCPD asymptotically achieves the Cram\'{e}r-Rao bound (CRB)\cite{kay} for parameter estimation. Furthermore, we provide an identifiability analysis establishing the uniqueness conditions of SCPD.

\item We propose an integrated algorithm for target tracking and adaptive beamforming. Target tracking is achieved by associating and fusing the parameters estimated via SCPD from different BS pairs. Two post-processing steps are introduced to enhance tracking robustness against outliers. The proposed algorithm incorporates the tracking result from the previous snapshot to assist the current estimation. Based on the tracking result at the current snapshot, an adaptive beamforming scheme is designed for the next snapshot to direct the transmitter toward the predicted target directions.
\end{itemize}
\par \emph{Notations}---A tensor, matrix, vector, and scalar are denoted by a bold calligraphic
letter $\pmb{\mathcal{X}}$, bold uppercase letter $\bm{X}$, bold lowercase letter $\bm{x}$, and regular letter $x$ (or $X$), respectively. In particular, $\bm{I}_n$, $\bm{0}_{n\times m}$, and $\bm{1}_{n}$ denote the $n\times n$ identity matrix, the $n\times m$ zero matrix, and the length-$n$ all-ones vector, respectively. The $i$th entry of a vector $\bm{x}$ is denoted by $[\bm{x}]_i$. For a matrix $\bm{X}$, its $(i,j)$th entry, $i$th row, and $j$th column are represented as $[\bm{X}]_{i,j}$, $[\bm{X}]_{i,:}$, and $[\bm{X}]_{:,j}$, respectively. $\overline{\bm{X}}$ and $\underline{\bm{X}}$ are obtained from $\bm{X}$ by removing its first and last rows, respectively. The $(i,j,k)$th entry and the $(j,k)$th mode-1 fiber of a third-order tensor $\pmb{\mathcal{Y}}$ are denoted by $[\pmb{\mathcal{Y}}]_{i,j,k}$ and $[\pmb{\mathcal{Y}}]_{:,j,k}$; mode-2 and mode-3 fibers are defined analogously. The mode-$i$ unfolding of $\pmb{\mathcal{Y}}$ is represented as $\bm{Y}^{(i)}$, and $\times_i$ denotes the mode-$i$ product, $\forall i\in \{1,2,3\}$\cite{modei}. Let $\mathbb{R}$, $\mathbb{C}$, and $\mathbb{N}$ denote the real, complex, and positive integer spaces, respectively. Superscripts $ (\cdot)^T$, $(\cdot)^*$, $(\cdot)^H$, $(\cdot)^{-1}$, and $(\cdot)^{\dagger}$ denote the transpose, conjugate, Hermitian transpose, inverse, and pseudo-inverse operators, respectively. Symbols $\mathcal{R}(\cdot)$, $\mathcal{I}(\cdot)$, $\mathcal{M}(\cdot)$, $\mathcal{V}(\cdot)$, $\llbracket \cdot \rrbracket$, $\otimes$, $\odot$, $\circledast$, and $\oslash$ denote the real part, imaginary part, mean, variance, Kruskal operator, Kronecker product, Khatri-Rao product, Hadamard product, and element-wise division, respectively. For any complex vector $\bm{x}$ composed of multiple sinusoids, the operator $\angle(\bm{x})$ extracts the frequency parameter of the dominant sinusoid using the matrix pencil method \cite{sarkar1995using}. For any complex matrix $\bm{X}$ whose $l$th column is a single-tone sinusoid, the operator $\angle_l(\bm{X})$ extracts the frequency parameter of $[\bm{X}]_{:,l}$ using the principal-singular-vector utilization for modal analysis (PUMA) method\cite{puma}. As a geometry-related expression in the Cartesian coordinate system, the operator $\angle_{\mathrm{s}}(\bm{x}_1,\bm{x}_2)$ returns the signed broadside angle from a 2D location vector $\bm{x}_1\in \mathbb{R}^2$ to another location vector $\bm{x}_2\in \mathbb{R}^2$.
\section{Signal Model}
\label{sec:signal_model}
Consider a networked ISAC system with $D$ BSs, yielding a total of $J \triangleq D(D-1)/2$ BS pairs. Each BS is equipped with a uniform linear array with $M$ transmit and receive antennas. In the downlink communication, each BS transmits $K$ OFDM pilot symbols over $N$ subcarriers to multiple UEs it serves. These pilot signals are simultaneously reused as sensing waveforms to illuminate surrounding targets. To mitigate inter-cell interference, we follow the assumption in \cite{5GNR} that different BSs transmit signals over disjoint bandwidth parts (BWPs). In the meanwhile, each BS can receive signals over all active BWPs in the system and thus can separate transmissions from different BSs through BWP indexes.
\begin{figure}[!t]
\centerline{\includegraphics[width=8.5cm]{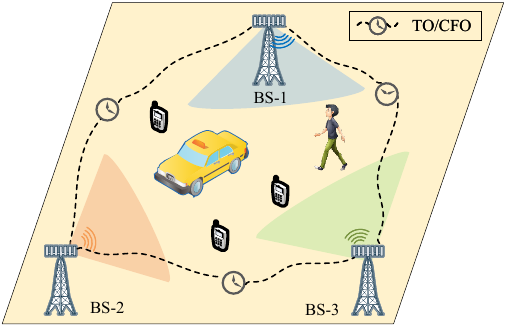}}
\centering
\caption{A networked ISAC system with multi-BSs collaboratively performing communication and sensing. TO and CFO exist among BSs, which must be compensated for reliable sensing.}
\label{times}
\end{figure}
\par Assume that the $j$th BS pair consists of two BSs, namely BS-$j_1$ and BS-$j_2$, where $j\in \{1,\cdots, J\}$, $j_1\in \{1,\cdots, D\}$, and $j_2\in \{1,\cdots, D\}$. In other words, $j_1$ and $j_2$ denote the indexes of the first and second BSs in the $j$th pair, respectively. Let $\bm{y}_{j_1,n,k}$ (and $\bm{y}_{j_2,n,k})$ denote the signal transmitted from BS-$j_2$ to targets to BS-$j_1$ (and from BS-$j_1$ to targets to BS-$j_2$) at the $k$th OFDM symbol and the $n$th subcarrier. It can be modeled as
\begin{equation}
\bm{y}_{i,n,k}\in \mathbb{C}^M=\bm{H}_{i,n,k}\cdot \bm{w}_i \cdot s_{i,n,k} +\bm{z}_{i,n,k}, \ \forall i\in \{j_1,j_2\},
\label{rece}
\end{equation} 
where $\bm{z}_{i,n,k}\in \mathbb{C}^M$ represents the noise. $s_{j_1,n,k}\in \mathbb{C}$ (and $s_{j_2,n,k}\in \mathbb{C}$) denotes the $k$th OFDM symbol transmitted from BS-$j_2$ (and BS-$j_1$) on the $k$th subcarrier. $\bm{w}_{j_1}\in \mathbb{C}^M$ (and $\bm{w}_{j_2}\in \mathbb{C}^M$) denotes the transmit beamformer at BS-$j_2$ (and BS-$j_1$). $\bm{H}_{j_1,n,k}$ (and $\bm{H}_{j_2,n,k}$) denotes the sensing channel transmitted from BS-$j_2$ to targets to BS-$j_1$ (and from BS-$j_1$ to targets to BS-$j_2$), modeled as\cite{sensingModel}
\begin{equation}
\begin{aligned}
\bm{H}_{i,n,k}  & \in \mathbb{C}^{M\times M} = \sum_{l=1}^L \alpha_{i,l} \cdot e^{-j2\pi \Delta f (n-1)(\tau_{j,l}+\Delta \tau_i)}\\
& \!\!\!\!\!\!\!\! \cdot e^{j2\pi T_{\text{sym}}(k-1) (\nu_{j,l}+\Delta \nu_i)}
\cdot \bm{a}_{\mathrm{r}}(\theta_{i,l})\cdot \bm{a}_{\mathrm{t}}^T(\phi_{i,l}), \forall i\in \{j_1,j_2\},
\end{aligned}
\label{chan}
\end{equation}
where $\Delta f$ denotes the subcarrier spacing, and $T_{\text{sym}}$ represents the OFDM symbol duration. $\Delta \tau_{j_1}$ and $\Delta \nu_{j_1}$ are TO and CFO, respectively, in the BS-$j_2$ to targets to BS-$j_1$ link. Similarly, $\Delta \tau_{j_2}$ and $\Delta \nu_{j_2}$ are TO and CFO in the BS-$j_1$ to targets to BS-$j_2$ link. $\alpha_{i,l}$, $\phi_{i,l}$, $\theta_{i,l}$, $\tau_{j,l}$, and $\nu_{j,l}$ represent the reflection coefficient, angle-of-departure (AoD), angle-of-arrival (AoA), bistatic delay, and Doppler shift of the $l$th target at the $j$th BS pair, respectively. Specifically, the delay parameter is modeled based on the bistatic geometry as
\begin{equation}
\tau_{j,l}=c^{-1}\cdot (\|\bm{u}_{l}-\bm{u}_{j_1}\|_2+\|\bm{u}_{l}-\bm{u}_{j_2}\|_2),
\label{range}
\end{equation}
where $c$ denotes the speed of light. $\bm{u}_l\in \mathbb{R}^2$, $\bm{u}_{j_1}\in \mathbb{R}^2$, and
$\bm{u}_{j_2}\in \mathbb{R}^2$ represent 2D locations of the $l$th target, BS-$j_1$, and BS-$j_2$ in the Cartesian coordinate system. The receive steering vector $\bm{a}_{\mathrm{r}}(\theta_{i,l})$ is expressed as
\begin{equation}
\bm{a}_{\mathrm{r}}(\theta_{i,l}) \in \mathbb{C}^{M}=[1, e^{j\frac{2\pi}{\lambda}d \sin(\theta_{i,l})}, \ldots, e^{j\frac{2\pi}{\lambda}d (M-1)\sin(\theta_{i,l})}]^T,
\label{mknewnew}
\end{equation}
where $d$ and $\lambda$ denote the antenna spacing and carrier wavelength, respectively. The transmit steering vector $\bm{a}_{\mathrm{t}}(\phi_{i,l})$ can be defined in a similar fashion to \eqref{mknewnew}.
\par Of particular note is the bistatic reciprocity property \cite{offset,radar} inherent in the channel model \eqref{chan}. It states that TOs and CFOs in the two links at the $j$th BS pair are reciprocal as\cite{offset}
\begin{equation}
\Delta \tau_{j_1}=-\Delta \tau_{j_2}, \quad \Delta \nu_{j_1}=-\Delta \nu_{j_2}, \ \forall j\in \{1,\cdots,J\}.
\label{or_prop}
\end{equation}
Furthermore, the two links at the $j$th BS pair share the same delay $\tau_{j,l}$ and Doppler shift $\nu_{j,l}$ for each $l$th target. The angle parameters of the $l$th target at the $j$th BS pair are also reciprocal as $\theta_{j_1,l}=\phi_{j_2,l}$ and $\theta_{j_2,l}=\phi_{j_1,l}$, $\forall j,l$. The bistatic reciprocity will be efficiently used to estimate offset and target parameters in Sections \ref{SOEMPM} and \ref{sync}.
\section{SOE-MP and Extension}
\label{SOEMPM}
The SOE-MP method \cite{offset} was originally proposed for TO and CFO estimation in the single-input single-output (SISO) scenario. This section first reviews SOE-MP and then extends it to the MIMO scenario.
\subsection{Review of SOE-MP}
By reducing the models in \eqref{rece} and \eqref{chan} from MIMO to SISO scenarios, the received signals over $N$ subcarriers and $K$ OFDM symbols, denoted as $\bm{Y}_i\in \mathbb{C}^{N\times K}$, are expressed as 
\begin{equation}
\begin{aligned}
& [\bm{Y}_i]_{n,k}= \sum_{l=1}^L s_{i,n,k} \cdot \alpha_{i,l} \cdot e^{-j2\pi \Delta f (n-1)(\tau_{j,l}+\Delta \tau_i)} \\
& \ \ \ \ \ \ \ \, \cdot e^{j2\pi T_{\text{sym}}(k-1) (\nu_{j,l}+\Delta \nu_i)} +z_{i,n,k}, \ \forall i\in \{j_1,j_2\}, \ \forall n, k.
\end{aligned}
\label{reduced}
\end{equation} 
After removing all transmit pilots at the receiver, the delay-Doppler spectrum of received signals can be constructed as
\begin{equation}
\bm{E}_i\in\mathbb{C}^{N\times K}=(\bm{Y}_i\oslash \bm{S}_i) \times_1 \bm{F}_{N}^{-1} \times_2 \bm{F}_{K}, \ \forall i\in \{j_1,j_2\},
\label{eq:soemp_dd_spectrum}
\end{equation}
where $\bm{S}_i\in \mathbb{C}^{N\times K}$ with $[\bm{S}_i]_{n,k}=s_{i,n,k}$ for all $n$ and $k$. $\bm{F}_{N}$ and $\bm{F}_{K}$ denote the $N$-point and $K$-point DFT matrices. Then, SOE-MP selects one column and one row, respectively, in $\bm{E}_i$ with the maximum energy as\cite{offset}
\begin{equation}
k_i= {\arg \max \limits_{k}} \,\big\| [\bm{E}_i]_{:,k} \big\|_2^2, \quad n_i= {\arg \max \limits_{n}} \,\big\|[\bm{E}_i]_{n,:}\big\|_2^2.
\label{eq:soemp_max_indices}
\end{equation}
Accordingly, the matrix $\bm{E}_i$ is compressed along the time and frequency domains, respectively, into vectors as 
\begin{equation}
\bm c_{\tau,i}=\bm{F}_{N}\cdot [\bm{E}_i]_{:,k_i},
\ \, \bm c_{\nu,i}=\bm{F}_{K}^{-1} \cdot [\bm{E}^T_i]_{:,n_i}, \ \forall i\in \{j_1,j_2\}.
\label{eq:soemp_decoupled_vectors}
\end{equation}
Subsequently, SOE-MP leverages the bistatic reciprocity to match the compressed vectors in the two links of the $j$th BS pair as
\begin{equation}
\bm{d}_{\tau,j}\!\in\!\mathbb{C}^{N}\!=\!\bm{c}_{\tau,j_1}\circledast\bm{c}_{\tau,j_2}^{*},\ \ 
\bm{d}_{\nu,j}\!\in\!\mathbb{C}^{K}\!=\!\bm{c}_{\nu,j_1}\circledast\bm{c}_{\nu,j_2}^*. 
\label{eq:soemp_matched_vectors}
\end{equation}
\par Note that in the noiseless scenario, i.e., $z_{i,n,k}=0$ in \eqref{reduced}, both $\bm{d}_{\tau,j}$ and $\bm{d}_{\nu,j}$ are the superpositions of $(L^2-L+1)$ paths with distinct frequency parameters. Among these paths, the desired one has a frequency parameter determined by the TO or CFO of the $j$th BS pair, and may have the largest magnitude\cite{syncletter,offset}. Accordingly, SOE-MP applies the matrix pencil method \cite{sarkar1995using} to extract the frequency parameter of the dominant path in $\bm{d}_{\tau,j}$ and $\bm{d}_{\nu,j}$ for estimating the TO and CFO of the $j$th BS pair as 
\begin{equation}
{\Delta\hat{\tau}}_{j_2}= -\frac {\angle(\bm{d}_{\tau,j})}{4\pi\Delta f}, \quad {\Delta\hat{\nu}}_{j_2}=\frac{\angle(\bm{d}_{\nu,j})}{4\pi T_{\mathrm{sym}}}.
\label{eq:soemp_offsets}
\end{equation}
However, it is noteworthy that the path with the largest magnitude in $\bm{d}_{\tau,j}$ (and $\bm{d}_{\nu,j}$) is not always the desired one, due to the randomness of the magnitudes of all paths\cite{syncletter}. In such an undesired case, the TO (and CFO) estimated by SOE-MP corresponds to the frequency parameter of an interfering path rather than the desired path, leading to significant performance degradation of SOE-MP.
\subsection{Extension to MIMO Scenarios}
\label{secSOEMPM}
To extend SOE-MP from SISO to MIMO scenarios, we stack the received
signals $\bm{y}_{i,n,k}$ in \eqref{rece} over all $n$ and $k$ to form a
tensor $\pmb{\mathcal{Y}}_i\in \mathbb{C}^{M\times N\times K}$, where 
\begin{equation}[\pmb{\mathcal{Y}}_i]_{:,n,k}=\bm{y}_{i,n,k}, \quad \forall i\in \{j_1,j_2\}, \ \forall n, k.
\end{equation}
Then, we construct an angle-delay-Doppler spectrum similarly to \eqref{eq:soemp_dd_spectrum} as 
\begin{equation}
\pmb{\mathcal{E}}_i\in\mathbb{C}^{M\times N\times K}\!=\!
(\pmb{\mathcal{Y}}_i \oslash \pmb{\mathcal{S}}_i) \times_1 \bm{F}_{M}
\times_2 \bm{F}_{N}^{-1}
\times_3 \bm{F}_{K}, \ \forall i \! \in \! \{j_1,j_2\}, 
\label{eq:soempm_transform}
\end{equation}
where $\pmb{\mathcal{S}}_i\in \mathbb{C}^{M\times N\times K}$ with $[\pmb{\mathcal{S}}_i]_{:,n,k}=s_{i,n,k}$ for all $n$ and $k$. This spectrum is then compressed into two vectors as
\begin{equation}
\bm{c}_{\tau,i}=\bm{F}_{N}\cdot [\pmb{\mathcal{E}}_i]_{m_i,:,k_i},
\quad
\bm{c}_{\nu,i}=\bm{F}_{K}^{-1}\cdot [\pmb{\mathcal{E}}_i]_{a_i,b_i,:}, \ \forall i \in \{j_1,j_2\},
\label{eq:soempm_ref_vectors}
\end{equation}
where the indexes in $\pmb{\mathcal{E}}_i$ are selected similarly to \eqref{eq:soemp_max_indices} as
\begin{equation}
\left\{\begin{aligned}
& (m_i,k_i) =
{\arg \max \limits_{(m,k)}}\,\big\|[\pmb{\mathcal{E}}_i]_{m,:,k}\big\|_2^2,\\
& (a_i,b_i) =
{\arg \max \limits_{(a,b)}}\,\big\|[\pmb{\mathcal{E}}_i]_{a,b,:}\big\|_2^2.
\end{aligned}\right.
\label{mk}
\end{equation}
\par Based on $\bm{c}_{\tau,i}$ and $\bm{c}_{\nu,i}$ in \eqref{eq:soempm_ref_vectors}, TO and CFO can be estimated using the same matching and matrix pencil steps as in \eqref{eq:soemp_matched_vectors}
and \eqref{eq:soemp_offsets}. In addition to offset estimation, we further extend SOE-MP to estimate target parameters $\{\tau_{j,l},\nu_{j,l},
\theta_{j_1,l},\theta_{j_2,l}\}_{l=1}^L$ in MIMO scenarios, as detailed in Appendix \ref{app:soempm_sic}.
\section{SCPD and Analysis}
\label{sync}
To address the inter-path interference problem of SOE-MP, this section proposes a tensor-based estimator that enables the separation of multipath components from the received sensing signals. Specifically, the received signal tensor $\pmb{\mathcal{Y}}_i\oslash \pmb{\mathcal{S}}_i$ in \eqref{eq:soempm_transform} admits a CPD:
\begin{equation}
\pmb{\mathcal{Y}}_i\oslash \pmb{\mathcal{S}}_i = \llbracket  \bm{A}_i, \bm{B}_i, \bm{C}_i\rrbracket + \pmb{\mathcal{Z}}_i\oslash \pmb{\mathcal{S}}_i,\ \forall i\in \{j_1,j_2\},
\label{eq1}
\end{equation}
where the factor matrices $\bm{A}_i\in \mathbb{C}^{M \times L}$, $\bm{B}_i\in \mathbb{C}^{N \times
L}$, and $\bm{C}_i\in \mathbb{C}^{K \times L}$ are formulated as
\begin{equation}
\!\!\left\{\begin{aligned}
& \!\,[\bm{A}_i]_{:,l}= \alpha_{i,l}\cdot\boldsymbol{a}^T_{\mathrm{t}}(\phi_{i,l})\cdot \boldsymbol
w_i\cdot \boldsymbol{a}_{\mathrm{r}}(\theta_{i,l}), \\
& \!\,[\bm{B}_i]_{:,l}\!=\![1, e^{-j2\pi \Delta f (\tau_{j,l}+\Delta \tau_i)}, \ldots, e^{-j2\pi (N-1) \Delta f (\tau_{j,l}+\Delta \tau_i)}]^T\!, 
\\
& \!\,[\bm{C}_i]_{:,l}\!=\![1, e^{j2\pi T_{\text{sym}}(\nu_{j,l}+\Delta \nu_i)}, \ldots, e^{j2\pi (K-1)T_{\text{sym}} (\nu_{j,l}+\Delta \nu_i)}]^T.
\end{aligned}\right.
\label{SCPD_offset_recovery}
\end{equation}
These three factor matrices can be estimated from the noise-corrupted tensor $\pmb{\mathcal{Y}}_i\oslash \pmb{\mathcal{S}}_i$ via an approximate low-rank CPD, then used to estimate offset and target parameters. Notably, we incorporate the a priori Vandermonde structure of factor matrices to enhance the approximation accuracy of CPD. This enhancement is achieved by enforcing structural constraints in CPD, leading to the SCPD method:
\begin{equation}
\begin{aligned}
& \mathop {\min}\limits_{ \bm{A}_i,\bm{B}_i, \bm{C}_i}
\big\| \llbracket \bm{A}_i, \bm{B}_i, \bm{C}_i\rrbracket - \pmb{\mathcal{Y}}_i\oslash \pmb{\mathcal{S}}_i\big\|_F^2 \\
& \ \ \ \text{s.t.} \ \bm{A}_i   \in  \mathbb{S}, \ \bm{B}_i   \in  \mathbb{S}, \ \bm{C}_i   \in  \mathbb{S},
\end{aligned}
\label{q10}
\end{equation}
where $\mathbb{S}$ denotes the set of Vandermonde matrices up to column scaling as
\begin{equation}
\begin{aligned}
\mathbb{S} \triangleq & \Big\{ \bm{X}\in \mathbb{C}^{Q \times L}: \ [\bm{X}]_{:,l} = \lambda_l\cdot \bm{v}_Q(\omega_l), \, \exists\lambda_l \in \mathbb{C}, \\
& \ \ \, \exists \omega_l \in [0,2\pi), \forall l\in \{1,\cdots, L \}, \ \forall Q \in \mathbb{N}, \, \forall L \in \mathbb{N} \Big\},
\end{aligned}
\label{set}
\end{equation}
and $\bm{v}_Q(\omega_l)$ denotes the Vandermonde vector parameterized by the generator $\omega_l$ as 
\begin{equation}
\bm{v}_Q(\omega_l)\in \mathbb{C}^{Q \times 1}\!=\![1, e^{-j2\pi \omega_l}, e^{-j2\pi \cdot 2\omega_l}, \ldots, e^{-j2\pi \cdot (Q-1)\omega_l}]^{T}.
\label{steering}
\end{equation}
Additionally, the scaling coefficient $\lambda_l$ is introduced in \eqref{set} to  accommodate the scaling ambiguity \cite{zhou,RZ,chen2026tensor,syncletter,Vandermonde,qian2018tensor} in CPD.
\subsection{Optimization}
\par Problem \eqref{q10} can be solved using the CALS framework within $R$ iterations, as summarized in Algorithm
\ref{alg1}. The $r$th iteration of Algorithm \ref{alg1} consists of three
subproblems for updating $\bm{A}^{r+1}_i$, $\bm{B}^{r+1}_i$, and $\bm{C}_i^{r+1}$
in turn. These three subproblems can be expressed in a similar fashion to
\begin{equation}
\min \limits_{ \bm{X}\in \mathbb{C}^{Q \times L}} \| \bm{D}\bm{X}^T - \bm{E}\|_F^2, \ \text {s.t. } \bm{X}\in \mathbb{S},
\label{subthree}
\end{equation}
whose optimal solution is challenging to obtain due to its nonconvex constraint. Alternatively, we obtain an approximate yet efficient solution to
\eqref{subthree} in two steps. The first step ignores the constraint $\bm{X}\in \mathbb{S}$ and solves \eqref{stage1} via least-squares (LS) to yield $\bar{\bm{X}}$. The second step
takes the constraint into account to refine $\bar{\bm{X}}$ via \eqref{stage2}:
\begin{numcases}{}
\bar{\bm{X}} = \arg \min \limits_{ \bm{X}} \| \bm{D}\bm{X}^T -
\bm{E}\|_F^2, \label{stage1} \\
\bm{X}= \arg \min \limits_{ \bm{X}} \big\|  \bar{\bm{X}} -
\bm{X}\big\|_F^2, \ \text {s.t. } \bm{X}\in \mathbb{S}.
\label{stage2}
\end{numcases}
Problem \eqref{stage2} can be efficiently solved through a correlation-based
scheme\cite{zhou} as
\begin{equation}
{[\bm{X}]}_{:,l}= \mu_{l} \cdot \bm{v}_Q(\omega_l), \ \forall l\in \{1,\cdots, L \},
\label{subBfinal}
\end{equation}
where $\mu_{l}=\bm{v}^{\dagger}_Q(\omega_l)\cdot [\bar{\bm{X}}]_{:,l}$ is obtained
from LS, and $\omega_l$ from 
\begin{equation}
\omega_l= \arg \max \limits_{\omega} \frac{\big|[\bar{\bm{X}}]^H_{:,l}\cdot \bm{v}_Q(\omega)\big|}{\big\|[\bar{\bm{X}}]_{:,l}\big\|_2\cdot \|\bm{v}_Q(\omega)\|_2}.
\label{subBend}
\end{equation}
It is noteworthy that \eqref{subBfinal} constitutes a maximum likelihood
estimator under the assumption that the entries of the estimation error matrix (relative to the true matrix $\bm{X}$) are independently and identically drawn from a circularly symmetric Gaussian distribution\cite{zhou}.
\par After obtaining the matrices
$\{\hat{\bm{A}}_i,\hat{\bm{B}}_i,\hat{\bm{C}}_i\}, \forall i\in \{j_1,j_2\}$, via Algorithm \ref{alg1}, we incorporate the signal model in \eqref{SCPD_offset_recovery} with the bistatic reciprocity to estimate the offset and target parameters relative to the $j$th BS pair as
\begin{equation}
\left\{\begin{aligned}
& \Delta \hat{\tau}_{j_2} =  \frac{\mathcal{M}( \hat{\bm{b}}_{j_1} - \hat{\bm{b}}_{j_2})}{4\pi \Delta f}, \ \ 
 \Delta \hat{\nu}_{j_2} =  \frac{\mathcal{M}( \hat{\bm{c}}_{j_2} - \hat{\bm{c}}_{j_1})}{4\pi T_{\text{sym}}}, \\
& \hat{\tau}_{j,l} = -\frac{[\hat{\bm{b}}_{j_1}]_l + [\hat{\bm{b}}_{j_2}(\bm{o})]_l}{4\pi \Delta f}, \ \
\hat{\nu}_{j,l} = \frac{[\hat{\bm{c}}_{j_1}]_l + [\hat{\bm{c}}_{j_2}(\bm{o})]_l}{4\pi T_{\text{sym}}}, \ \forall l,\\
& \hat{\theta}_{j_1,l} = \arcsin \! \Bigg(\!\frac{\lambda\! \cdot \! [\hat{\bm{a}}_{j_1}]_l}{2\pi d}\!\Bigg), \  
 \hat{\theta}_{j_2,l} = \arcsin \! \Bigg(\!\frac{\lambda\! \cdot \! [\hat{\bm{a}}_{j_2}(\bm{o})]_l}{2\pi d}\!\Bigg), \forall l,
\end{aligned}\right.
\label{mk}
\end{equation}
where $\hat{\bm{a}}_i \in \mathbb{R}^{L} = \big[\angle_1(\hat{\bm{A}}_i),\cdots, \angle_L(\hat{\bm{A}}_i)\big]^T$, and $\hat{\bm{b}}_i$ and $\hat{\bm{c}}_i$ are defined similarly, $\forall i\in \{j_1,j_2\}$. In \eqref{mk}, $\bm{o}\in \mathbb{N}^L$ denotes a permutation vector obtained through data association between the two links of the $j$th BS pair as
\begin{equation}
\begin{aligned}
& \min_{\bm{o}\in\mathbb{O}}
\Big\{{\mathcal{V}\big(\hat{\bm{b}}_{j_1} - \hat{\bm{b}}_{j_2}(\bm{o})- 4\pi\Delta f \cdot \Delta\hat{\tau}_{j_2}\cdot \bm{1}_L\big)}\\
& \ \ \ \ \ \ \ + {\mathcal{V}\big(\hat{\bm{c}}_{j_1}  -  \hat{\bm{c}}_{j_2}(\bm{o})+ 4\pi T_{\mathrm{sym}} \cdot\Delta\hat{\nu}_{j_2}\cdot \bm{1}_L \big)} \Big\},
\end{aligned}
\label{eq:asso}
\end{equation}
where the function $\bm{x}(\bm{o})\in \mathbb{R}^L$ for any $\bm{x}\in \mathbb{R}^L$ is defined as $[\bm{x}(\bm{o})]_l=[\bm{x}]_{[\bm{o}]_l}$, $\forall l$. The feasible set $\mathbb{O}$ contains $L!$ vectors, where the first vector is $[1,\cdots,L]^T$ and all other vectors are formed by permuting the entries of the first vector. As an explanation, the objective function value of \eqref{eq:asso} equals zero when the estimates $\Delta\hat{\tau}_{j_2}$, $\Delta\hat{\nu}_{j_2}$, $\hat{\bm{b}}_i$, and $\hat{\bm{c}}_{i}$, $\forall i\in \{j_1,j_2\}$, equal their true values. In addition to the estimates in \eqref{mk}, the other parameters relative to the $j$th BS pair can be estimated based on the bistatic reciprocity as $\Delta \hat{\tau}_{j_1}=-\Delta \hat{\tau}_{j_2}$, $\Delta \hat{\nu}_{j_1}=-\Delta \hat{\nu}_{j_2}$, $\hat{\phi}_{j_1,l}=\hat{\theta}_{j_2,l}$ and $\hat{\phi}_{j_2,l}=\hat{\theta}_{j_1,l}$, $\forall l$.
\begin{algorithm}[!htbp]
\caption{Solve \eqref{q10} via CALS}
\begin{algorithmic}[1]
\REQUIRE $\pmb{\mathcal{Y}}_i \oslash \pmb{\mathcal{S}}_i\in \mathbb{C}^{M\times N\times K}$, $L$, $R$, and $\delta$. 
\STATE {\textbf{Initialize}} $ \bm{B}^1_i$ and $
        {\bm{C}}^1_i$. \FOR{$r=1,\cdots, R$} \STATE Update $\bm{A}^{r+1}_i$ via
        \eqref{subthree}, where $\bm{X} \triangleq \bm{A}^{r+1}_i$, $\bm{D}
        \triangleq \bm{B}_i^r\odot  \bm{C}_i^r$, and $\bm{E} \triangleq
        \bm{Y}_i^{(1)} \oslash \bm{S}_i^{(1)}$. \STATE Update  $\bm{B}^{r+1}_i$ via \eqref{subthree},
        where $\bm{X} \triangleq \bm{B}^{r+1}_i $, $\bm{D} \triangleq
        \bm{C}_i^r\odot  \bm{A}_i^{r+1}$, and $\bm{E} \triangleq
        \bm{Y}_i^{(2)} \oslash \bm{S}_i^{(2)}$. \STATE Update  $\bm{C}^{r+1}_i$ via \eqref{subthree},
        where $\bm{X} \triangleq \bm{C}^{r+1}_i $, $\bm{D} \triangleq
        \bm{A}_i^{r+1}\odot  \bm{B}_i^{r+1}$, and $\bm{E} \triangleq
        \bm{Y}_i^{(3)} \oslash \bm{S}_i^{(3)}$. \STATE Check the convergence condition
        $|\epsilon_{r+1}-\epsilon_r|/\epsilon_r < \delta$, where
        $\epsilon_{r+1}=\big\|\llbracket \bm{A}^{r+1}_i,\bm{B}^{r+1}_i,
        \bm{C}^{r+1}_i \rrbracket - \pmb{\mathcal{Y}}_i \oslash \pmb{\mathcal{S}}_i \big\|_F$. \ENDFOR \ENSURE
        $\hat{\bm{A}}_i = \bm{A}^{r+1}_i$, $\hat{\bm{B}}_i = \bm{B}^{r+1}_i$, and $\hat{\bm{C}}_i = \bm{C}^{r+1}_i$.
\end{algorithmic}
\label{alg1}
\end{algorithm}
\subsection{Uniqueness Conditions}
This section analyzes the uniqueness conditions for SCPD. A CPD in \eqref{eq1} is deemed unique if all factor matrices therein are identifiable up to scaling and permutation ambiguities in the noiseless scenario where $\pmb{\mathcal{Z}}_i=\bm{0}$\cite{zhou,RZ,chen2026tensor,syncletter,Vandermonde,qian2018tensor}. Beyond the standard CPD in \eqref{eq1}, the proposed SCPD method in \eqref{q10} further exploits the structural constraints $\bm{A}_i\in  \mathbb{S}$, $\bm{B}_i \in \mathbb{S}$, and $\bm{C}_i \in \mathbb{S}$. By taking these constraints into account, we provide the uniqueness conditions for SCPD as follows.
\par \emph{Lemma 1:} Let $\pmb{\mathcal{Y}}_i\oslash \pmb{\mathcal{S}}_i \in \mathbb{C}^{M\times
N\times K} =  \llbracket  \bm{A}_i, \bm{B}_i,
\bm{C}_i\rrbracket$ with $\bm{A}_i \in \mathbb{S}$, $\bm{B}_i \in \mathbb{S}$, and $\bm{C}_i \in \mathbb{S}$. Suppose that the columns in $\bm{A}_i$ (and likewise $\bm{B}_i$ and $\bm{C}_i$) have distinct generators, i.e., $\omega_t\neq \omega_l$, $\forall t \neq l$ in the definition of $\mathbb{S}$ in \eqref{set}. If there exist integer pairs $\{P_t,Q_t\}_{t=1}^3$ satisfying $(P_1+Q_1)=M+1$, $(P_2+Q_2)=N+1$, $(P_3+Q_3)=K+1$, and
\begin{equation}
\left\{\begin{array}{l}
{\text{rank}}\big([\bm{A}_i]_{1:P_1-1,:} \odot [\bm{B}_i]_{1:P_2,:}\odot [\bm{C}_i]_{1:P_3,:}\big)=L,\\
{\text{rank}}\big([\bm{A}_i]_{1:Q_1,:}\odot [\bm{B}_i]_{1:Q_2,:}\odot [\bm{C}_i]_{1:Q_3,:}\big)=L,
\end{array}\right.
\label{kjss_shift}
\end{equation}
then $\pmb{\mathcal{Y}}_i\oslash \pmb{\mathcal{S}}_i$ admits a unique rank-$L$ CPD.
\par \emph{Proof:} Given $\bm{A}_i \in \mathbb{S}$, $\bm{B}_i \in \mathbb{S}$, and $\bm{C}_i \in \mathbb{S}$, the spatial smoothing technique \cite{Vandermonde} can be applied to the tensor $\pmb{\mathcal{Y}}_i\oslash \pmb{\mathcal{S}}_i$. Specifically, the spatial smoothing operator $\mathcal{F}(\pmb{\mathcal{X}})$ for any $\pmb{\mathcal{X}} \in \mathbb{C}^{M\times N\times K}$ is defined under parameters $\{P_t,Q_t\}_{t=1}^3$ as
\begin{equation}
\begin{aligned}
& \mathcal{F}(\pmb{\mathcal{X}}) \in \mathbb{C}^{(P_1P_2P_3)\times (Q_1Q_2Q_3)} \triangleq \big[  
\bm{J}_{1,1,1} \cdot \bm{x}, \bm{J}_{1,1,2} \cdot  \bm{x},\cdots,\\
& \ \bm{J}_{1,1, Q_3} \cdot  \bm{x}, \bm{J}_{1,2,1} \cdot \bm{x}, \cdots, \bm{J}_{Q_1,Q_2,(Q_3-1)} \cdot \bm{x}, \bm{J}_{Q_1,Q_2,Q_3} \cdot \bm{x}\big],
\end{aligned}
\label{unfolding}
\end{equation}
where $\bm{x} = {\text{vec}}(\pmb{\mathcal{X}})$ and $
\bm{J}_{q_1, q_2, q_3} \triangleq  \bm{J}_{q_1}  \otimes \bm{J}_{q_2} \otimes \bm{J}_{q_3}, \forall q_1, q_2, q_3$, with $\bm{J}_{q_t} \triangleq [\bm{0}_{P_t \times(q_t-1)} , \bm{I}_{P_t},
\bm{0}_{P_t \times(Q_t-q_t)}]$, $\forall t\in \{1,2,3\}$. By the definition of spatial smoothing, it can be verified that
\begin{equation}
\begin{aligned}
\mathcal{F}(\pmb{\mathcal{Y}}_i\oslash \pmb{\mathcal{S}}_i) = \, &   [\tilde{\bm{A}}_i]_{1:P_1,:}\odot [\bm{B}_i]_{1:P_2,:}\odot [\bm{C}_i]_{1:P_3,:} \cdot  \bm{\Lambda}_1 \\
& \cdot \big([\tilde{\bm{A}}_i]_{1:Q_1,:}\odot [\bm{B}_i]_{1:Q_2,:}\odot [\bm{C}_i]_{1:Q_3,:}\big)^T,
\end{aligned}
\label{folding}
\end{equation}
where $\bm{\Lambda}_1=\text{diag}\big(\alpha_{i,1}\cdot \boldsymbol{a}^T_{\mathrm{t}}(\phi_{i,1})\cdot\boldsymbol
w_i, \ldots, \alpha_{i,L}\cdot\boldsymbol{a}^T_{\mathrm{t}}(\phi_{i,L})\cdot\boldsymbol
w_i\big)$ is derived from \eqref{SCPD_offset_recovery}, and $\tilde{\bm{A}}_i=\bm{A}_i\bm{\Lambda}_1^{-1}$.
\par Let $\mathcal{F}(\pmb{\mathcal{Y}}_i\oslash \pmb{\mathcal{S}}_i)=\boldsymbol{U}\bm{\Sigma}\boldsymbol{V}^H$ be the compact singular value decomposition (SVD) of the matrix $\mathcal{F}(\pmb{\mathcal{Y}}_i\oslash \pmb{\mathcal{S}}_i)$. Due to \eqref{kjss_shift} and \eqref{folding}, there exists a nonsingular matrix $\boldsymbol{O} \in \mathbb{C}^{L \times L}$
such that \cite{Vandermonde}
\begin{equation} 
\boldsymbol{U}\boldsymbol{O}=[\tilde{\bm{A}}_i]_{1:P_1,:}\odot [\bm{B}_i]_{1:P_2,:}\odot [\bm{C}_i]_{1:P_3,:}.
\label{anglepara}
\end{equation}
Furthermore, the Vandermonde structure of $\tilde{\boldsymbol{A}}$ implies that\cite{Vandermonde}
\begin{equation} 
\boldsymbol{U}_2 \boldsymbol{O}=\boldsymbol{U}_1 \boldsymbol{O}\bm{\Lambda}_2,
\label{anglepara2}
\end{equation}
where $\bm{\Lambda}_2=\text{diag}\big(e^{j2 \pi d \lambda^{-1} \sin (\theta_{i,1})}, \ldots, e^{j{2 \pi d \lambda^{-1}\sin (\theta_{i,L})}}\big)$ and 
\begin{equation}
\left\{\begin{array}{l}
\bm{U}_1=\big( [\bm{I}_{(P_1-1)},\bm{0}_{(P_1-1)\times 1}]^T \odot \bm{I}_{P_2}\odot \bm{I}_{P_3}\big)\cdot  \boldsymbol{U}, \\
\bm{U}_2= \big( [\bm{0}_{(P_1-1)\times 1}, \bm{I}_{(P_1-1)}] \odot \bm{I}_{P_2}\odot \bm{I}_{P_3}\big)\cdot  \boldsymbol{U}.
\end{array}\right.
\end{equation}
Based on the eigenvalue decomposition (EVD) of
$\boldsymbol{U}_1^{\dagger} \boldsymbol{U}_2=\boldsymbol{O} \bm{\Lambda}_2
\boldsymbol{O}^{-1}$, one can estimate the angle parameters
$\{\theta_{i,l}\}_{l=1}^L$ exactly up to permutation ambiguity from the matrix $\bm{\Lambda}_2$. Then, the matrices $\bm{A}_i$ and $\tilde{\bm{A}}_i$ can be exactly recovered based on $\{\theta_{i,l}\}_{l=1}^L$ up to scaling and permutation ambiguities. 
\par By combining the recovered matrix $\tilde{\bm{A}}_i$ with \eqref{anglepara}, we can obtain the first $P_2$ and $P_3$ rows of $\bm{B}_i$ and $\bm{C}_i$, respectively, as
\begin{equation}
\left\{\begin{array}{l}
{[\bm{B}_i]}_{1:P_2,l}= \bm{I}_{P_2} \otimes [1,\bm{0}_{1\times(P_3-1)}] \cdot \bm{a}_l/\|\bm{a}_l\|_2, \ \forall l,
\\{[\bm{C}_i]}_{1:P_3,l} = [1,\bm{0}_{1\times(P_2-1)}] \otimes  \bm{I}_{P_3} \cdot \bm{a}_l/\|\bm{a}_l\|_2, \ \forall l,
\end{array}\right.
\label{end1}
\end{equation}
where $\bm{a}_l \triangleq \big( [\tilde{\bm{A}}_i]^H_{:,l} \otimes \bm{I}_{(P_2P_3)}\big) \cdot  \bm{U} \cdot [\bm{O}]_{:,l}$ is used for simplicity. Moreover, due to the Vandermonde structures of $\bm{B}_i$ and $\bm{C}_i$, their generators defined in \eqref{SCPD_offset_recovery} can be identified from \eqref{end1} as 
\begin{equation}
\left\{\begin{array}{l}
e^{-j2\pi \Delta f (\tau_{j,l}+\Delta \tau_i)}=[\bm{B}_i]_{1:P_2-1,l}^{\dagger}\cdot [\bm{B}_i]_{2:P_2,l}, \ \forall l, \\
e^{j2\pi T_{\text{sym}}(\nu_{j,l}+\Delta \nu_i)}=[\bm{C}_i]_{1:P_3-1,l}^{\dagger}\cdot [\bm{C}_i]_{2:P_3,l}, \ \forall l.
\end{array}\right.
\end{equation}
\par In conclusion, all factor matrices $\bm{A}_i$, $\bm{B}_i$, and $\bm{C}_i$ in CPD are identifiable up to scaling and permutation ambiguities, and thus $\pmb{\mathcal{Y}}_i\oslash \pmb{\mathcal{S}}_i$ admits a unique rank-$L$ CPD. $\hfill\blacksquare$
\subsection{Cram\'{e}r--Rao Bound}
We analyze the performance limit of SCPD by deriving the CRB for parameter estimation. Given the measurements $\pmb{\mathcal{Y}}_i \in \mathbb{C}^{M\times N\times K}$ and $\pmb{\mathcal{S}}_i \in \mathbb{C}^{M\times N\times K}$ for $i\in \{j_1,j_2\}$ at the $j$th BS pair, let $\bm{\Omega}  \in \mathbb{R}^{8L+2} = [\bm{\zeta}, \bm{\xi}]^T $ denote the vector of all unknown parameters, where $\bm{\zeta} =
[\Delta\tau_{j_2}, \Delta\nu_{j_2}]$, $\bm{\xi}\in \mathbb{R}^{8L}=[\bm{\xi}_{1}, \ldots, \bm{\xi}_{L}]$, and $\bm{\xi}_{l}\in \mathbb{R}^{8}$ contains the parameters related to the $l$th target as
\begin{equation}
\!\bm{\xi}_{l} \!=\! [\tau_{j,l}, \nu_{j,l}, \theta_{j_2,l}, \phi_{j_2,l}, \mathcal{R}(\alpha_{j_1,l}), \mathcal{I}(\alpha_{j_1,l}), \mathcal{R}(\alpha_{j_2,l}), \mathcal{I}(\alpha_{j_2,l})].
\end{equation}
Due to the bistatic reciprocity, as shown in \eqref{or_prop}, the parameters $\Delta \tau_{j_1}$, $\Delta \nu_{j_1}$, $\theta_{j_1,l}$, and $\phi_{j_1,l}$ are not involved in $\bm{\Omega}$ since they are dependent on $\Delta \tau_{j_2}$, $\Delta \nu_{j_2}$, $\phi_{j_2,l}$, and $\theta_{j_2,l}$, respectively. 
\par The Fisher information matrix (FIM) with respect to $\bm{\Omega}$ based on the received signals $\bm{y}_{i,n,k}$ in \eqref{rece}, $\forall i,n,k$, can be constructed as
\begin{equation}
\bm{F} \in
\mathbb{R}^{(8L+2)\times (8L+2)} =
\begin{bmatrix}
\bm{F}_{\bm{\zeta}\bm{\zeta}} & \!\!\! \bm{F}_{\bm{\zeta}\bm{\xi}} \\
\bm{F}_{\bm{\xi}\bm{\zeta}} & \!\!\! \bm{F}_{\bm{\xi}\bm{\xi}}
\end{bmatrix},
\label{FIM}
\end{equation}
where $\bm{F}_{\bm{\zeta}\bm{\zeta}} \in
\mathbb{R}^{2\times 2}$ and
$\bm{F}_{\bm{\xi}\bm{\xi}} \in \mathbb{R}^{8L \times 8L}$ characterize the information on the offset and target parameters, respectively, and $\bm{F}_{\bm{\zeta}\bm{\xi}} =
\bm{F}_{\bm{\xi}\bm{\zeta}}^T \in \mathbb{R}^{2\times 8L}$ characterizes the cross-information between $\bm{\zeta}$ and $\bm{\xi}$.
Suppose that in \eqref{rece}, $\bm{z}_{i,n,k}$ is an independent circularly symmetric complex Gaussian noise vector with zero mean and variance $\sigma_i^2$ for $i\in\{j_1,j_2\}$. Then, the $(p, q)$th entry of the matrix $\bm{F}$ can be calculated based on the Slepian structure \cite{kay} as
\begin{equation}
\begin{aligned}
[\bm{F}]_{p,q} &= \frac{2}{\sigma_{j_2}^2} \sum_{n=1}^N \sum_{k=1}^K \mathcal{R} \left(\left( \frac{\partial \bm{\mu}_{j_2,n,k}}{\partial ([\bm{\Omega}]_p)} \right)^H \left( \frac{\partial \bm{\mu}_{j_2,n,k}}{\partial ([\bm{\Omega}]_q)} \right) \right) \\
&\quad \, + \frac{2}{\sigma_{j_1}^2} \sum_{n=1}^N \sum_{k=1}^K \mathcal{R} \left( \left( \frac{\partial \bm{\mu}_{j_1,n,k}}{\partial ([\bm{\Omega}]_p)} \right)^H \left( \frac{\partial \bm{\mu}_{j_1,n,k}}{\partial ([\bm{\Omega}]_q)} \right) \right),
\end{aligned}
\label{fim_eq}
\end{equation}
where $\bm{\mu}_{i,n,k} = \sum_{l=1}^L
\alpha_{i,l}\cdot \bm{b}_{i,n,k,l}$ for $i\in \{j_1,j_2\}$ is defined with 
\begin{equation}
\begin{aligned}
\bm{b}_{i,n,k,l} = \ & e^{-j2\pi \Delta f (n-1)(\tau_{j,l}+\Delta \tau_i)} \cdot e^{j2\pi T_{\text{sym}}(k-1)(\nu_{j,l}+\Delta \nu_i)} \\
& \cdot \bm{a}_{\text{r}}(\theta_{i,l}) \cdot \bm{a}_{\text{t}}^T(\phi_{i,l})\cdot  \bm{w}_i \cdot s_{i,n,k}.
\end{aligned}
\end{equation}
Specifically, the partial derivatives of $\bm{\mu}_{i,n,k}$ for $i\in \{j_1,j_2\}$ with respect to each element of $\bm{\Omega}$ are given by
\begin{equation}
\left\{\begin{aligned}
& \frac{\partial \bm{\mu}_{i,n,k}}{\partial \Delta \tau_{j_2}} =  j 2\pi \Delta f(n-1)\cdot e_i \cdot \bm{\mu}_{i,n,k}, \\
& \frac{\partial \bm{\mu}_{i,n,k}}{\partial \Delta \nu_{j_2}} = -j 2\pi T_{\text{sym}}(k-1) \cdot e_i \cdot \bm{\mu}_{i,n,k},\\
& \frac{\partial \bm{\mu}_{i,n,k}}{\partial \tau_{j,l}} = -j 2\pi \Delta f(n-1) \cdot \alpha_{i,l} \cdot \bm{b}_{i,n,k,l},\\
& \frac{\partial \bm{\mu}_{i,n,k}}{\partial \nu_{j,l}} = j 2\pi T_{\text{sym}}(k-1) \cdot  \alpha_{i,l} \cdot \bm{b}_{i,n,k,l},\\
& \frac{\partial \bm{\mu}_{i,n,k}}{\partial \mathcal{R}(\alpha_{i,l})} = \bm{b}_{i,n,k,l}, \ \ \frac{\partial \bm{\mu}_{i,n,k}}{\partial \mathcal{I}(\alpha_{i,l})} = j\cdot \bm{b}_{i,n,k,l}, \\
& \frac{\partial \bm{\mu}_{j_1,n,k}}{\partial \phi_{j_2,l}} = \bm{g}_{j_1,k,n,l}, \ \ \frac{\partial \bm{\mu}_{j_2,n,k}}{\partial \theta_{j_2,l}} = \bm{g}_{j_2,k,n,l},\\
& \frac{\partial \bm{\mu}_{j_1,n,k}}{\partial \theta_{j_2,l}} =\bm{h}_{j_1,k,n,l}, \ \  \frac{\partial \bm{\mu}_{j_2,n,k}}{\partial \phi_{j_2,l}} = \bm{h}_{j_2,k,n,l},
\end{aligned}\right.
\label{deri}
\end{equation}
where $e_{j_1}=1$ and $e_{j_2}=-1$. $\bm{g}_{i,n,k,l}\in \mathbb{C}^{M}$ and $\bm{h}_{i,n,k,l}\in \mathbb{C}^{M}$ for $i\in \{j_1,j_2\}$ are defined as
\begin{equation}
\left\{\begin{aligned}
\bm{g}_{i,n,k,l}& =\alpha_{i,l} \cdot \bm{b}_{i,n,k,l} \circledast ( j{2\pi d}\lambda^{-1} \cos(\theta_{i,l}) \cdot \bm{m} ) , \\
\bm{h}_{i,n,k,l} &= \alpha_{i,l} \cdot \big(\bm{a}_{\text{t}}(\phi_{i,l}) \circledast (j{2\pi d} \lambda^{-1} \cos(\phi_{i,l})\cdot \bm{m}) \big)^T \!\cdot \bm{w}_i \\
& \!\!\!\!\!\!\!\!\!\!\!\!\!\!\!\!\!\!\cdot e^{-j2\pi \Delta f (n-1)(\tau_{j,l}+\Delta \tau_i)} e^{j2\pi T_{\text{sym}}(k-1)(\nu_{j,l}+\Delta \nu_i)} \bm{a}_{\text{r}}(\theta_{i,l}) s_{i,n,k},
\end{aligned}\right.
\label{deri2}
\end{equation}
where $\bm{m}\in \mathbb{R}^M = [0, 1, \ldots, M-1]^T$.
After calculating the FIM in \eqref{FIM}, we obtain the CRB for offset estimation in \eqref{offsetCRB} and target parameter estimation in \eqref{targetCRB} via the Schur complement\cite{kay} as 
\begin{numcases}{}
\text{CRB}(\bm{\zeta}) = \big( \bm{F}_{\bm{\zeta}\bm{\zeta}} - \bm{F}_{\bm{\zeta}\bm{\xi}} \cdot \bm{F}_{\bm{\xi}\bm{\xi}}^{-1} \cdot  \bm{F}_{\bm{\xi}\bm{\zeta}} \big)^{-1}, 
\label{offsetCRB} \\
\text{CRB}(\bm{\xi}) = \big( \bm{F}_{\bm{\xi}\bm{\xi}} - \bm{F}_{\bm{\xi}\bm{\zeta}} \cdot \bm{F}_{\bm{\zeta}\bm{\zeta}}^{-1} \cdot  \bm{F}_{\bm{\zeta}\bm{\xi}} \big)^{-1}. 
\label{targetCRB}
\end{numcases}
\section{Target Tracking with Adaptive Beamforming}
\label{NetworkLevel}
Considering the continuous operation of the system over time, this section tracks the 2D trajectories and 2D velocities of moving targets. An adaptive beamforming scheme is also developed to ensure tracking performance, as summarized in Algorithm \ref{alg:tracking_flow} and detailed below.  
\subsection{Adaptive Beamforming}
\label{beamdesign}
Before detailing the beamforming scheme, we first clarify the scenario extension from Sections \ref{sec:signal_model}-\ref{sync} to Section \ref{NetworkLevel} as follows. Assume that in a networked ISAC system, all BS pairs continuously collect $T$ snapshots of measurements $\{\pmb{\mathcal{Y}}_{j_a,t} \oslash \pmb{\mathcal{S}}_{j_a,t}\}_{t=1,j=1,a=1}^{T,J,2}$, and each measurement follows the signal model in \eqref{eq1}. Due to target movement, their locations and velocities vary across snapshots. At the first snapshot ($t=1$), random beamforming is applied to collect the measurements $\{\pmb{\mathcal{Y}}_{j_a,1} \oslash \pmb{\mathcal{S}}_{j_a,1}\}_{j=1,a=1}^{J,2}$, which are then used to estimate initial target locations as in \eqref{eq:network_joint_fusion2}. For all subsequent snapshots ($t>1$), we use the target locations estimated at the $t$th snapshot, denoted as $\{\hat{\bm{u}}_{l,t}\}_{l=1}^L$, to design adaptive beamformers for the $(t+1)$th snapshot, as detailed below. 
\par Based on $\{\hat{\bm{u}}_{l,t}\}_{l=1}^L$, the AoD of the $l$th target relative to the $d$th BS at the $t$th snapshot can be estimated as $\hat{\phi}_{d,l,t}=\angle_{\mathrm{s}}\big(\hat{\bm{u}}_{l,t},\bm{u}_d\big)$, where $\bm{u}_d\in \mathbb{R}^2$ denotes the 2D location of the $d$th BS, $\forall d\in \{1,\cdots, D\}$. Then, a transmit steering matrix for the $d$th BS at the $t$th snapshot is constructed as $\hat{\bm{A}}_{d,t}\in \mathbb{C}^{M\times L}$, where $[\hat{\bm{A}}_{d,t}]_{:,l}=\bm{a}_{\mathrm{t}}(\hat{\phi}_{d,l,t})$, $\forall l$. Compute the rank-1 truncated SVD of $\hat{\bm{A}}_{d,t}^*$ and denote the resulting left singular vector as $\bm{p}_{d,t}$. Accordingly, the adaptive beamformer for the $d$th BS at the $(t+1)$th snapshot, denoted as $\bm{w}_{d,t+1}$, is designed as
\begin{equation}
\bm{w}_{d,t+1}=\rho\cdot \frac{\bm{p}_{d,t}}{\|\bm{p}_{d,t}\|_2}, \quad \forall d\in \{1,\cdots, D\},\ t\in \{2,\cdots, T\},
\label{beam}
\end{equation}
where $\rho$ represents the transmit power budget. As an explanation, \eqref{beam} enables the system to concentrate transmit power toward the subspace spanned by the predicted target directions. 
\par After calculating \eqref{beam}, we apply a post-processing step, i.e., the \emph{consistency check} described in Section \ref{tracksub}, to enhance the robustness of the beamforming design against outliers. If the consistency check succeeds, the $d$th BS employs the beamformer in \eqref{beam} for the $(t+1)$th snapshot. Conversely, a failed consistency check indicates that the location estimates $\{\hat{\bm{u}}_{l,t}\}_{l=1}^L$ are unreliable outliers. In this case, the $d$th BS employs a random beamformer instead of the unreliable one in \eqref{beam} for the $(t+1)$th snapshot.
\subsection{Data Association}
\label{NetworkLevel_subA}
This subsection associates target parameters estimated at different BS pairs as a pre-processing step for tracking. From the measurements $\{\pmb{\mathcal{Y}}_{j_a,t} \oslash \pmb{\mathcal{S}}_{j_a,t}\}_{t=1,j=1,a=1}^{T,J,2}$, target parameters $
\mathcal{Z}_{j,t}\triangleq \{\hat{\tau}_{j,l,t},\hat{\nu}_{j,l,t},
\hat{\theta}_{j_1,l,t},\hat{\theta}_{j_2,l,t}\}_{l=1}^L$ are estimated independently for each $j$th BS pair and each $t$th snapshot, as described in Section \ref{sync}. Since the permutation of $L$ targets in the set $\{\mathcal{Z}_{j,t}\}_{j=1,t=1}^{J,T}$ may vary across $j$ and $t$, we propose a data association solution that treats the cases $t>1$ and $t=1$ differently. This different treatment arises because for $t>1$, data from the $(t-1)$th snapshot can be exploited to aid data association at the $t$th snapshot.
\par When $t=1$, data association is achieved by clustering localization results \cite{reid2003algorithm,liang2025robust} from all BS pairs. Before clustering, each BS pair independently locates $L$ targets as
\begin{equation}
\hat{\bm{d}}_{j,l}=
\arg \min \limits_{\bm{d}} f\big(\bm{d}\, \big|\,\mathcal{Z}_{j,1},l\big), \quad \forall j,l,
\label{eq:network_rough_pair_position}
\end{equation} 
where $\hat{\bm{d}}_{j,l}\in \mathbb{R}^2$ denotes the estimated 2D location of the $l$th target from the $j$th BS pair. The function $f$ is defined as
\begin{equation}
\begin{aligned}
& f(\bm{d}\, \big|\,\mathcal{Z}_{j,t},l) \triangleq 
\big(\|\bm{d}-\bm{u}_{j_1}\|_2 + \|\bm{d}-\bm{u}_{j_2}\|_2
-c\cdot \hat{\tau}_{j,l,t}\big)^2\\
& \ \ \ \ \ \ \, + \big(\angle_{\mathrm{s}}(\bm{d},\bm{u}_{j_1}) - 
\hat{\theta}_{j_1,l,t}\big)^2+\big(\angle_{\mathrm{s}}(\bm{d},\bm{u}_{j_2}) - 
\hat{\theta}_{j_2,l,t}\big)^2,\ \forall t,
\end{aligned}
\label{eq:network_pair_label_cost}
\end{equation}
where $\bm{u}_{j_1}$ and $\bm{u}_{j_2}$ are defined in \eqref{range}. In \eqref{eq:network_pair_label_cost}, the first term is derived from \eqref{range} to quantify the bistatic range fitting error, and the second and third terms quantify the angle fitting errors. Problem \eqref{eq:network_rough_pair_position} can be efficiently solved by first searching on a coarse grid of locations and then refining the search locally around the candidate point. After obtaining $\{\hat{\bm{d}}_{j,l}\}_{j=1,l=1}^{J,L}$ by solving \eqref{eq:network_rough_pair_position}, all these $JL$ location vectors are clustered into $L$ groups\cite{liang2025robust}. Each group contains $J$ elements, with its $j$th element being a vector selected from the set $\{\hat{\bm{d}}_{j,l}\}_{l=1}^L$, $\forall j$. Let $\hat{\bm{b}}_l\in \mathbb{R}^2$ denote the clustering center of the $l$th group, $\forall l$. Then, the data association for $t=1$ is achieved by 
\begin{equation}
\bm{o}_{j,1}=
\arg \min \limits_{\bm{o}\in \mathbb{O}} \sum\limits_{l=1}^{L}\tilde{f}
\big([\bm{o}]_l \, \big|\, \mathcal{Z}_{j,1},\hat{\bm{b}}_l\big), \quad \forall j, 
\label{eq:network_reference_pairing0}
\end{equation}
where $\bm{o}_{j,t}\in \mathbb{N}^L$ (with $t=1$ here) specifies the permutation of $L$ targets in $\mathcal{Z}_{j,t}$, and the feasible set $\mathbb{O}$ is defined in \eqref{eq:asso}. The function $\tilde{f}$ is defined similarly to the function $f$ as 
\begin{equation}
\tilde{f} \big([\bm{o}]_l \, \big|\,\mathcal{Z}_{j,t},\hat{\bm{b}}_l\big) \triangleq f \big(\hat{\bm{b}}_l\, \big|\, \mathcal{Z}_{j,t},[\bm{o}]_l\big), \quad \forall t.
\end{equation}
Problem \eqref{eq:network_reference_pairing0} can be solved via exhaustive search.
\par When $t>1$, we incorporate the localization result at the $(t-1)$th snapshot for data association at the $t$th snapshot. Let $\{\hat{\bm{u}}_{l,t-1}\}_{l=1}^L$ be target locations at the $(t-1)$th snapshot estimated via \eqref{eq:network_joint_fusion2}. Since the permutation of $L$ targets in $\mathcal{Z}_{j,t}$ may differ from that in $\{\hat{\bm{u}}_{l,t-1}\}_{l=1}^L$, we formulate an optimization problem similar to \eqref{eq:network_reference_pairing0} for data association: 
\begin{equation}
\bm{o}_{j,t}=
\arg \min \limits_{\bm{o}\in \mathbb{O}}
\sum\limits_{l=1}^{L}\tilde{f}
\big([\bm{o}]_l \, \big|\,\mathcal{Z}_{j,t},\hat{\bm{u}}_{l,t-1}\big), \ \forall j, \ \forall t\in \{2,\cdots, T\},
\label{eq:network_reference_pairing}
\end{equation}
which can be solved via exhaustive search.
\subsection{Data Fusion}
\label{tracksub}
Following data association, this subsection fuses the target parameter estimates $\{\mathcal{Z}_{j,t}\}_{j=1,t=1}^{J,T}$ across $j$ and $t$ to track their trajectories and velocities. Specifically, at the $t$th snapshot, the $l$th target is located through its angle and delay estimates as
\begin{equation}
\hat{\bm{u}}_{l,t}=\arg \min \limits_{\bm{u}} \sum\limits_{j=1}^J
g\big(\bm{u}\,\big| \,\mathcal{Z}_{j,t},[\bm{o}_{j,t}]_l\big), \quad \forall l,t,
\label{eq:network_joint_fusion}
\end{equation}
where the function $g$ is defined similarly to the function $f$ as
{\small{
\begin{equation}
g\big(\bm{u}\,\big| \, \mathcal{Z}_{j,t},[\bm{o}_{j,t}]_l\big)\!\triangleq \!\left\{\begin{array}{cl} \!\!\!\!\!\!\!\!\!\!\!\!\!\!\!\!\!\!\!\!\!\!\!\!\!\!\!\!\!\!\!\!\!\!\!\!\!\!\!\!\!\!\!
{0}, & \!\!\!\! \text{if }  {\hat{\tau}_{j,l,t}} \cdot  c\!<\! \|\bm{u}_{j_1}\!\!-\!\bm{u}_{j_2}\|_2; \\ 
\!\!\! f\big(\bm{u}\,\big| \, \mathcal{Z}_{j,t},[\bm{o}_{j,t}]_l\big), &  \!\!\!\! \text{otherwise.}
\end{array}\right.
\label{twoconditions}
\end{equation}}}To clarify, the first condition in \eqref{twoconditions} implies that the bistatic range estimate $(\hat{\tau}_{j,l,t}\cdot c)$ is unreliable since it is smaller than the distance between BS-$j_1$ and BS-$j_2$. Consequently, $\mathcal{Z}_{j,t}$ will not be used for fusion in \eqref{eq:network_joint_fusion}. To enhance tracking robustness against outliers, we refine the fusion result in \eqref{eq:network_joint_fusion} using two post-processing steps: \emph{outlier filtering} and \emph{consistency check}.
\par The \emph{outlier filtering} step stems from the fact that \eqref{eq:network_joint_fusion} yields a large objective value when the parameter estimate set $\{\mathcal{Z}_{j,t}\}_{j=1}^J$ contains outliers. Assume that $J_{\text{o}}$ out of $J$ elements in this set are outliers. We propose to filter them out by refining the objective function of \eqref{eq:network_joint_fusion}. Specifically, the sum over all $J$ elements in \eqref{eq:network_joint_fusion} is modified into the sum over only $(J-J_{\text{o}})$ elements as follows
\begin{equation}
\hat{\bm{u}}_{l,t}= \arg \min \limits_{\bm{u}} \Bigg(\mathop {\min} \limits_{c\in \{1,\cdots,C\}} \sum\limits_{j\in \mathbb{Q}_c}
g\big(\bm{u}\,\big| \,\mathcal{Z}_{j,t},[\bm{o}_{j,t}]_l\big)\Bigg), \ \forall l,t,
\label{eq:network_joint_fusion2}
\end{equation} 
where $C \triangleq \frac{J!}{(J-J_{\text{o}})!\cdot J_{\text{o}}!}$ is the number of combinations of selecting $(J-J_{\text{o}})$ elements from $J$, and $\mathbb{Q}_c\in \mathbb{N}^{J-J_{\text{o}}}$ denotes the element set for the $c$th combination, $\forall c\in \{1,\cdots,C\}$. For example, $\mathbb{Q}_1=\{1,2,\cdots,J-J_{\text{o}}\}$ consists of the first $(J-J_{\text{o}})$ elements. As an explanation, \eqref{eq:network_joint_fusion2} yields a significantly smaller objective value than \eqref{eq:network_joint_fusion} by eliminating the influence of $J_{\text{o}}$ outliers on the localization result.
\par The \emph{consistency check} step leverages the spatial consistency of a target between adjacent snapshots. Specifically, at each $t$th snapshot, the condition $\|\hat{\bm{u}}_{l,t}-\hat{\bm{u}}_{l,t-1}\|<\eta$ is checked for all $l\in \{1,\cdots,L\}$, where $\eta$ denotes a distance threshold. If this condition does not hold for a certain $l$, the displacement of the $l$th target between adjacent snapshots is too large to be physically unrealistic. Accordingly, the location estimates $\{\hat{\bm{u}}_{l,t}\}_{l=1}^L$ are deemed unreliable. In this case, they are replaced with the estimates from the previous snapshot, i.e., $\hat{\bm{u}}_{l,t} \triangleq \hat{\bm{u}}_{l,t-1}$, $\forall l$.
\par In addition to the trajectory, the 2D velocity of the $l$th target at the $t$th snapshot, denoted as $\bm{v}_{l,t}\in \mathbb{R}^2$, can also be tracked:
\begin{equation}
\hat{\bm{v}}_{l,t} =\lambda \cdot 
\bm{D}^{\dagger}(\hat{\bm{u}}_{l,t})\cdot \hat{\bm{\nu}}_{l,t}, \quad \forall l,t,
\label{eq:network_velocity_ls}
\end{equation}
where $\hat{\bm{\nu}}_{l,t}\in \mathbb{R}^J=[\hat{\nu}_{l,1,t},\cdots,\hat{\nu}_{l,J,t}]^T$ and the matrix $\bm{D}(\hat{\bm{u}}_{l,t})\in \mathbb{R}^{J\times 2}$ is defined as 
$$
[\bm{D}(\hat{\bm{u}}_{l,t})]_{j,:} \triangleq \frac{(\hat{\bm{u}}_{l,t}-\bm{u}_{j_1})^T}{\|\hat{\bm{u}}_{l,t}-\bm{u}_{j_1}\|_2}+\frac{(\hat{\bm{u}}_{l,t}-\bm{u}_{j_2})^T}{\|\hat{\bm{u}}_{l,t}-\bm{u}_{j_2}\|_2}, \quad \forall j.
$$ 
Note that \eqref{eq:network_velocity_ls} constitutes an LS estimator derived from the signal model $
\bm{D}(\bm{u}_{l,t})\cdot \bm{v}_{l,t} =
\lambda \cdot \bm{\nu}_{l,t}$, where $\hat{\bm{u}}_{l,t}$, $\hat{\bm{v}}_{l,t}$, and $\hat{\bm{\nu}}_{l,t}$ denote the estimates of the true parameters $\bm{u}_{l,t}$, $\bm{v}_{l,t}$, and $\bm{\nu}_{l,t}$, respectively.
\begin{algorithm}[!tbp]
\caption{Target Tracking with Adaptive Beamforming}
\begin{algorithmic}[1]
\REQUIRE $T$, $L$, $J_{\text{o}}$, $\eta$, $\rho$, and $\{\bm{u}_d\}_{d=1}^D$.
\STATE {\textbf{Initialize}} $\bm{w}_{d,1} \in \mathbb{C}^M$ randomly, $\forall d\in \{1,\cdots,D\}$.
\FOR{$t=1,\ldots,T$}
    \STATE Acquire the measurements $\{\pmb{\mathcal{Y}}_{j_a,t} \oslash \pmb{\mathcal{S}}_{j_a,t}\}_{j=1,a=1}^{J,2}$.
    \STATE Update $\{\mathcal{Z}_{j,t}\}_{j=1}^J$.
    \STATE Update $\{\bm{o}_{j,t}\}_{j=1}^J$ via \eqref{eq:network_reference_pairing0} or \eqref{eq:network_reference_pairing}.
    \STATE Update $\{\hat{\bm{u}}_{l,t}\}_{l=1}^L$ via \eqref{eq:network_joint_fusion2}.
    \STATE Update $\{\bm{w}_{d,t+1}\}_{d=1}^D$ randomly.
    \IF{$t>1$ \AND $\|\hat{\bm{u}}_{l,t}-\hat{\bm{u}}_{l,t-1}\|< \eta$ holds for all $l$,}
        \STATE Update $\{\bm{w}_{d,t+1}\}_{d=1}^D$ via \eqref{beam}.
    \ELSE 
        \STATE Update $\hat{\bm{u}}_{l,t} \triangleq \hat{\bm{u}}_{l,t-1}$, $\forall l\in \{1,\cdots,L\}$.
    \ENDIF
    \STATE Update $\{\hat{\bm{v}}_{l,t}\}_{l=1}^L$ via \eqref{eq:network_velocity_ls}.
\ENDFOR
\ENSURE $\{\hat{\bm{u}}_{l,t},\hat{\bm{v}}_{l,t}\}_{l=1,t=1}^{L,T}$.
\end{algorithmic}
\label{alg:tracking_flow}
\end{algorithm}
\begin{figure*}[!t]
\centering
{\includegraphics[width=3.5cm]{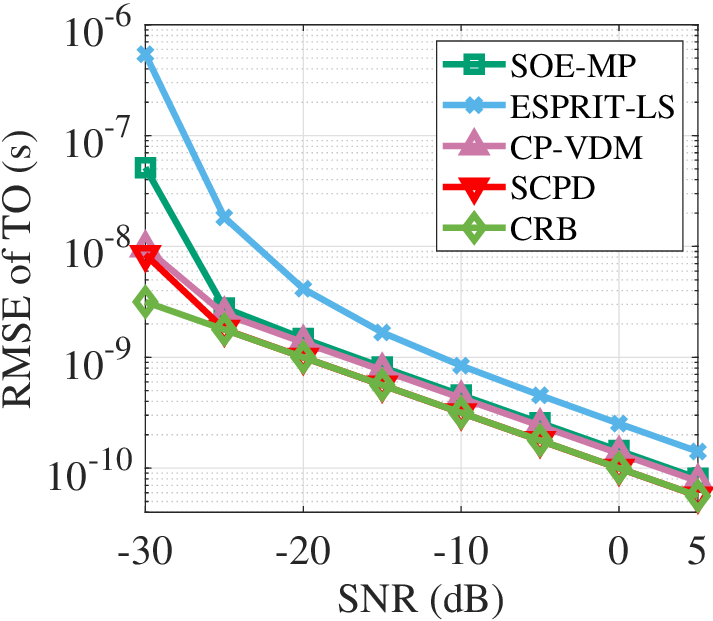}}
{\includegraphics[width=3.5cm]{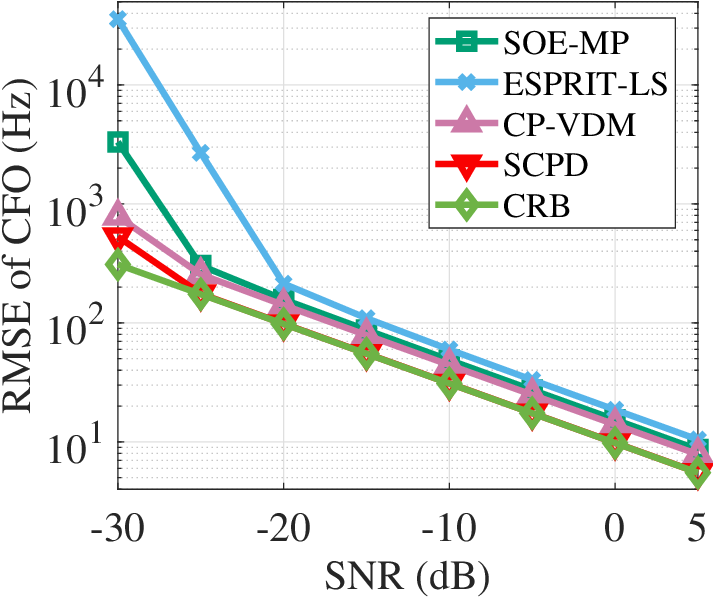}}
{\includegraphics[width=3.5cm]{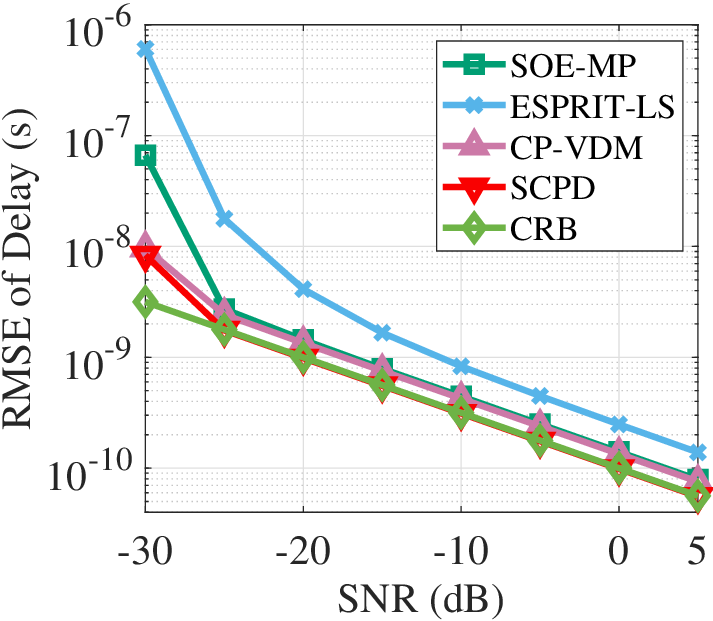}}
{\includegraphics[width=3.5cm]{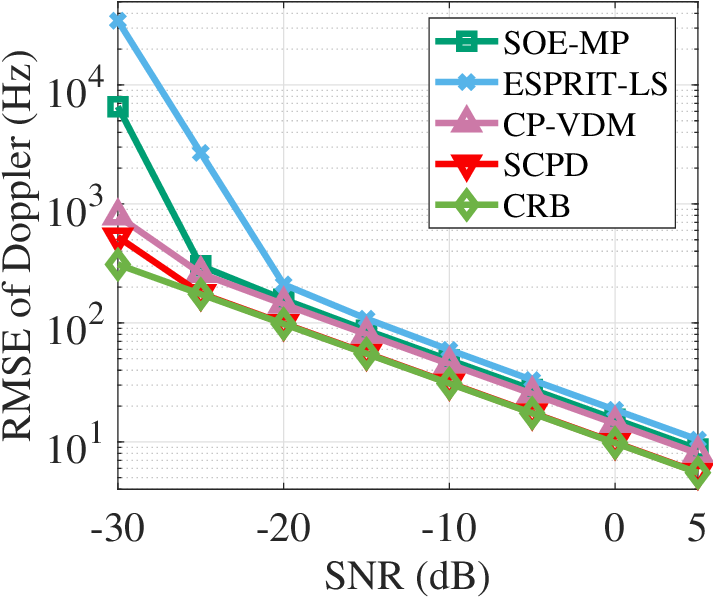}}
{\includegraphics[width=3.5cm]{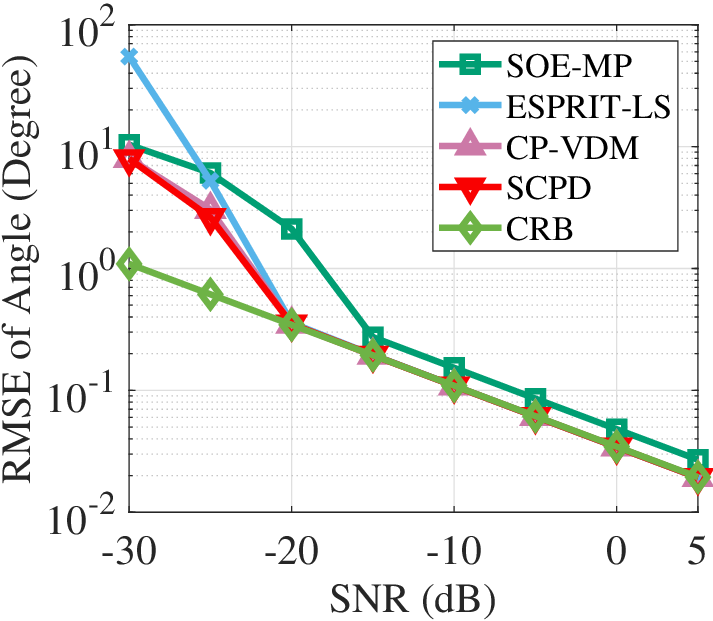}}
\caption{RMSE of parameter estimation versus SNR in the scenario where $L=1$.}
\label{L1_results}
\end{figure*}
\begin{figure*}[!t]
\centering
{\includegraphics[width=3.5cm]{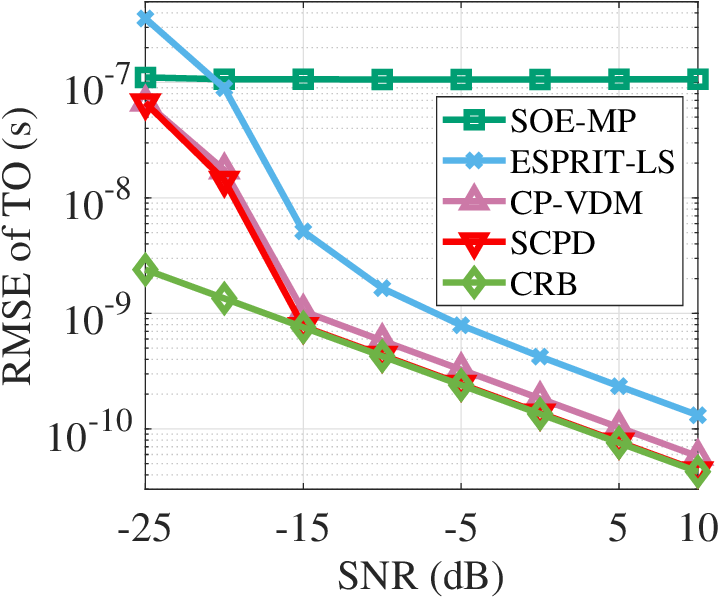}}
{\includegraphics[width=3.5cm]{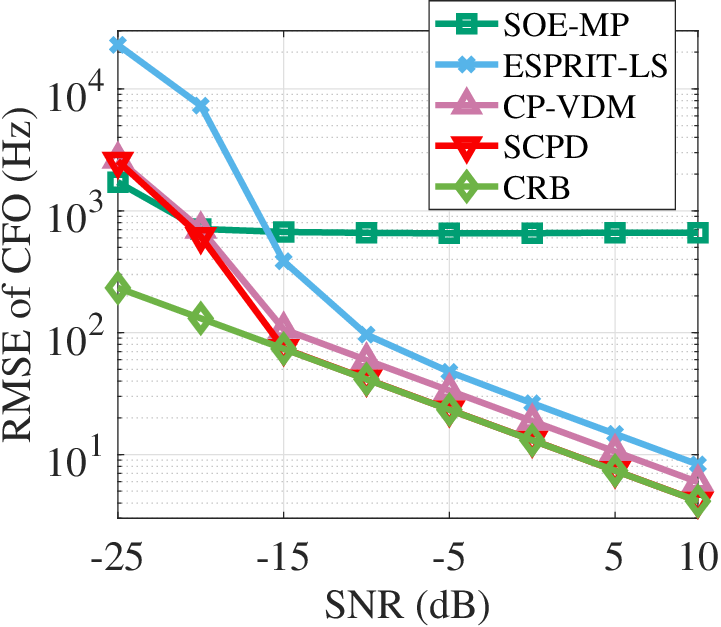}}
{\includegraphics[width=3.5cm]{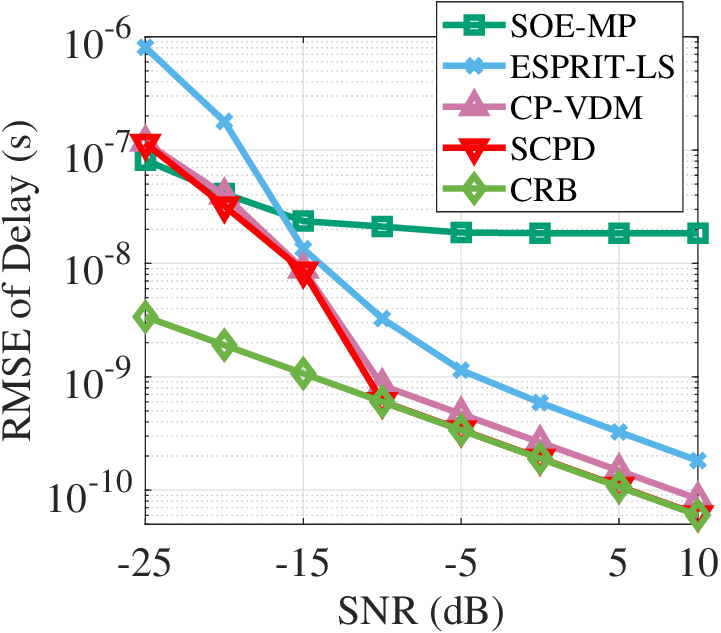}}
{\includegraphics[width=3.5cm]{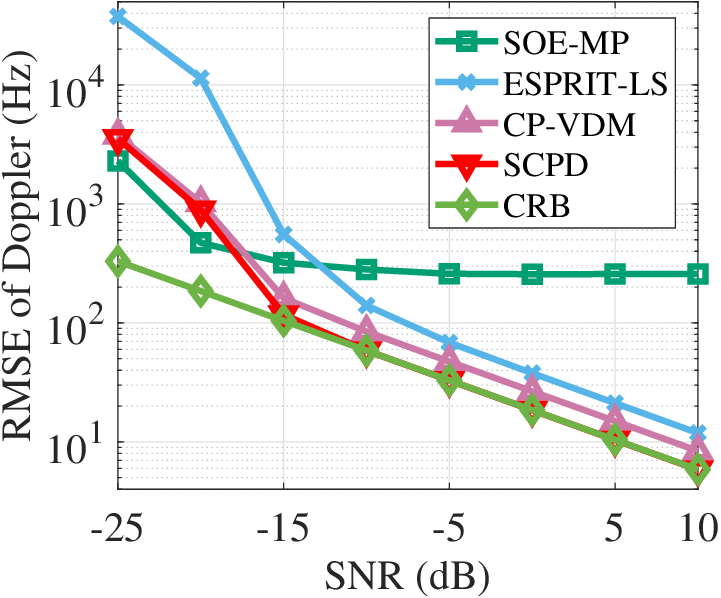}}
{\includegraphics[width=3.5cm]{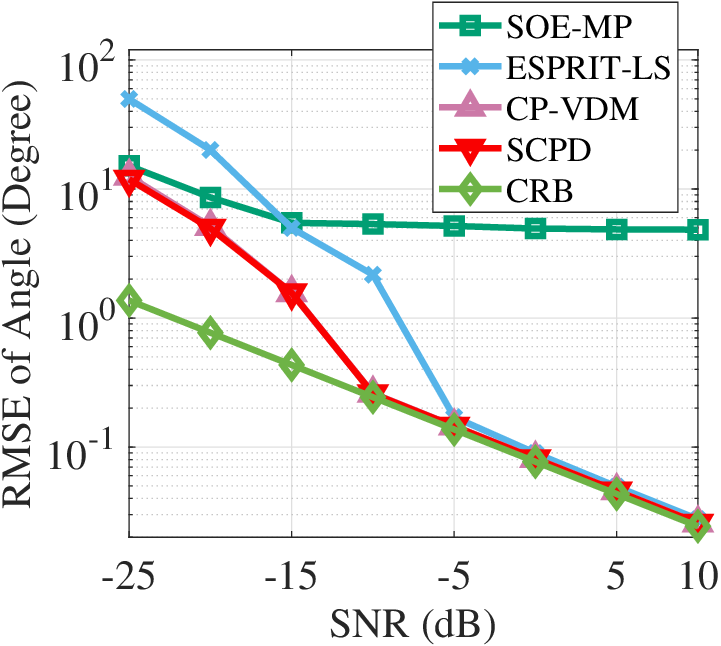}}
\caption{RMSE of parameter estimation versus SNR in the scenario where $L=2$.}
\label{L2_results}
\end{figure*}
\section{Simulation Results} 
This section validates the effectiveness of the proposed SCPD method for offset estimation and target sensing in networked ISAC. Its performance is compared with three baselines: SOE-MP\cite{offset}, ESPRIT-LS, and CP-VDM\cite{Vandermonde}. SOE-MP is a vector-based method as described in Section \ref{secSOEMPM} and Appendix \ref{app:soempm_sic}. ESPRIT-LS is a matrix-based method derived by combining the traditional ESPRIT\cite{xiang2023esprit} and LS, as detailed in Appendix \ref{sec:esprit_ls}. CP-VDM is a tensor-based method with an algebraic closed-form solution, whereas the proposed SCPD method is solved iteratively in Algorithm \ref{alg1}. We initialize SCPD using the result of CP-VDM. Specifically, the factor matrices $\bm{B}^1_i$ and ${\bm{C}}^1_i$, as shown in Line 1 of Algorithm \ref{alg1}, are initialized with their counterparts from CP-VDM. The parameter settings in Algorithms \ref{alg1} and \ref{alg:tracking_flow} include $R=20$, $\delta=10^{-4}$, $J_{\text{o}}=1$, $\eta=15  \, {\text{m}}$, and $\rho=1$. The simulation parameters include $M=10$, $N=36$, $K=20$, $d=\lambda/2$, $\lambda=c/f$ with a carrier frequency $f=28 \, {\text{GHz}}$, $\Delta f=B/(N-1)$ with a bandwidth $B=10 \, {\text{MHz}}$, and $T_{\text{sym}}=1.5/\Delta f$. The networked ISAC system comprises $D=4$ BSs located at the four corners of a square centered at the origin. Specifically, the coordinates of the 4 BSs are $\bm{u}_1=[80 \, {\text{m}},80 \, {\text{m}}]^T$, $\bm{u}_2=[-80 \, {\text{m}},-80 \, {\text{m}}]^T$, $\bm{u}_3=[80 \, {\text{m}},-80 \, {\text{m}}]^T$, and $\bm{u}_4=[-80 \, {\text{m}},80 \, {\text{m}}]^T$. Consequently, there are $J=6$ BS pairs in the system. The TO and CFO in each BS pair are randomly drawn from a normal distribution with zero mean and standard deviations of 10 ns and 100 Hz, respectively.
\subsection{Results on Parameter Estimation}
\label{paraest}
We evaluate the performance of different methods in estimating offset and target parameters, including $\{\Delta \tau_{j_2}, \Delta \nu_{j_2}, $ $\tau_{j,l}, \nu_{j,l}, \theta_{j_1,l},  \theta_{j_2,l}\}_{j=1,l=1}^{J,L}$, from the measurements $\{\pmb{\mathcal{Y}}_{j_a} \oslash \pmb{\mathcal{S}}_{j_a}\}_{j=1,a=1}^{J,2}$ at a single snapshot $(T=1)$. Estimation accuracy is measured using the root mean square error (RMSE). Specifically, for the $j$th BS pair, the RMSE of TO estimation is defined as ${\text{RMSE}}(\Delta \tau_{j_2}) = \sqrt{\mathbb{E} \big\{(\Delta \tau_{j_2}-\Delta \hat{\tau}_{j_2})^2\big\}}$, where $\mathbb{E}\{\cdot \}$ denotes the expectation over 500 Monte Carlo trials. The RMSEs of delay and angle estimation are respectively defined as ${\text{RMSE}}(\bm{\tau}_j) = \sqrt{\mathbb{E}
\big\{\frac{1}{L} \sum_{l=1}^L(\tau_{j,l}-\hat{\tau}_{j,l})^2\big\}}$ and ${\text{RMSE}}(\bm{\theta}_j)=\sqrt{\mathbb{E}
\big\{\frac{1}{2L} \sum_{l=1}^L\sum_{a=1}^2(\theta_{j_a,l}-\hat{\theta}_{j_a,l})^2\big\}}$. The RMSEs of CFO and Doppler shift estimation are defined similarly to those of TO and delay estimation, respectively. In each Monte Carlo trial, target locations are selected randomly and uniformly from the square surveillance area of side length $160$ m centered at the origin, and target speeds from the interval $[0,30\, \text{m/s}]$. Target directions are also selected uniformly at random.
\begin{table}[!t]
\centering
\caption{Success rates of parameter estimation, corresponding to the RMSE results in Figs. \ref{L1_results} and \ref{L2_results}. The best results in each SNR case are highlighted in bold, respectively.}
\label{tab:invalid_counts}
\setlength{\tabcolsep}{1.7mm}{
\begin{tabular}{|c|c|c|c|c|c|}
\hline
Targets & SNR (dB) & SOE-MP & ESPRIT-LS & CP-VDM & SCPD \\
\hline
\multirow{8}{*}{$L=1$}
& $-30$ & 95.27\% & 82.10\% & 97.30\% & \bf{97.87\%} \\
\cline{2-6}
& $-25$ & 98.03\% & 94.13\% & \bf{98.20\%} & \bf{98.20\%} \\
\cline{2-6}
& $-20$ & 98.37\% & 97.10\% & \bf{98.57\%} & \bf{98.57\%} \\
\cline{2-6}
& $-15$ & 98.77\% & 98.40\% & 98.87\% & \bf{99.00\%} \\
\cline{2-6}
& $-10$ & 98.90\% & 98.77\% & 99.03\% & \bf{99.23\%} \\
\cline{2-6}
& $-5$ & 99.33\% & 99.17\% & 99.27\% & \bf{99.50\%} \\
\cline{2-6}
& $0$ & 99.60\% & 99.37\% & 99.53\% & \bf{99.70\%} \\
\cline{2-6}
& $5$ & 99.60\% & 99.53\% & 99.67\% & \bf{99.70\%} \\
\hline
\multirow{8}{*}{$L=2$}
& $-25$ & \bf{94.97\%} & 64.13\% & 92.77\% & 93.73\% \\
\cline{2-6}
& $-20$ & \bf{96.57\%} & 86.10\% & 94.50\% & 95.33\% \\
\cline{2-6}
& $-15$ & \bf{97.37\%} & 94.63\% & 95.83\% & 96.63\% \\
\cline{2-6}
& $-10$ & \bf{97.87\%} & 96.23\% & 96.90\% & 97.40\% \\
\cline{2-6}
& $-5$ & \bf{97.97\%} & 96.33\% & 97.67\% & 97.87\% \\
\cline{2-6}
& $0$ & 98.00\% & 97.30\% & 98.07\% & \bf{98.37\%} \\
\cline{2-6}
& $5$ & 98.00\% & 98.17\% & 98.67\% & \bf{98.80\%} \\
\cline{2-6}
& $10$ & 98.13\% & 98.50\% & 99.03\% & \bf{99.13\%} \\
\hline
\end{tabular}}
\end{table}
\begin{figure*}[!t]
\centering
{\includegraphics[width=3.5cm]{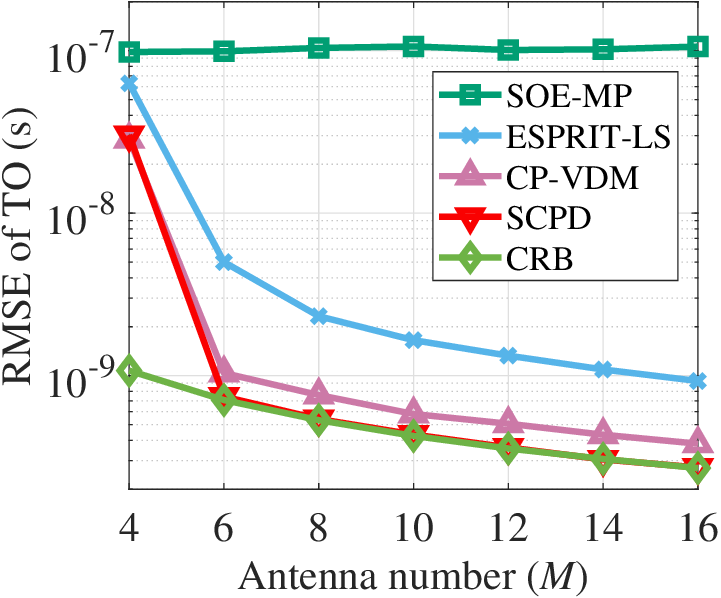}}
{\includegraphics[width=3.5cm]{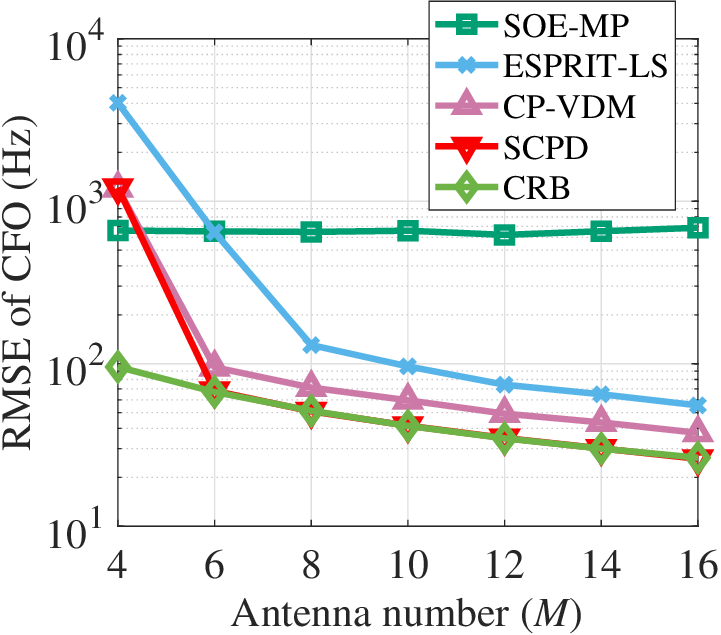}}
{\includegraphics[width=3.5cm]{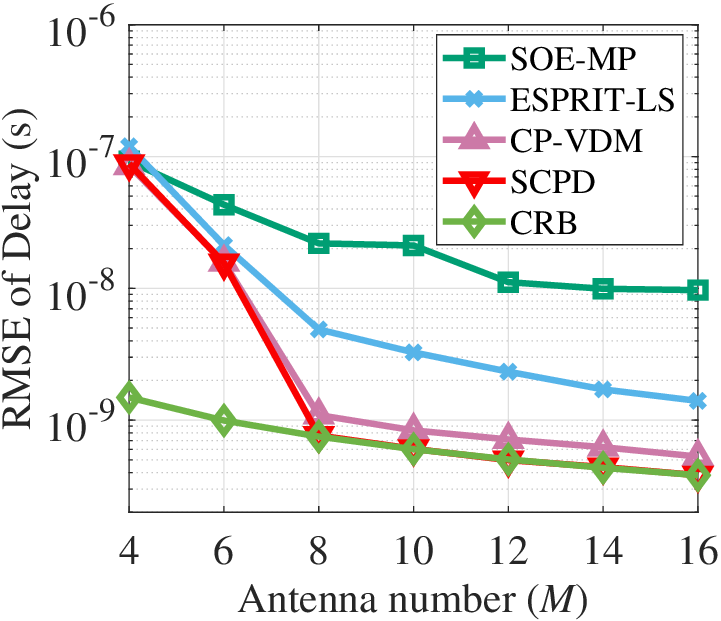}}
{\includegraphics[width=3.5cm]{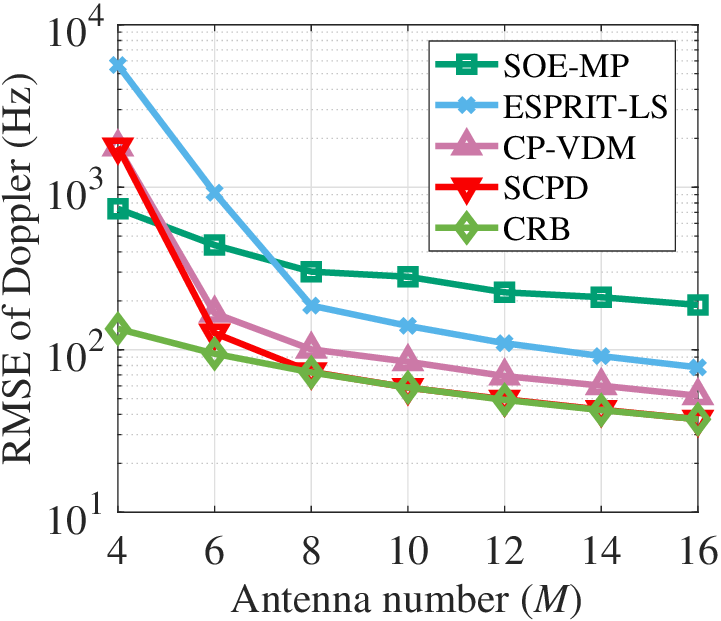}}
{\includegraphics[width=3.5cm]{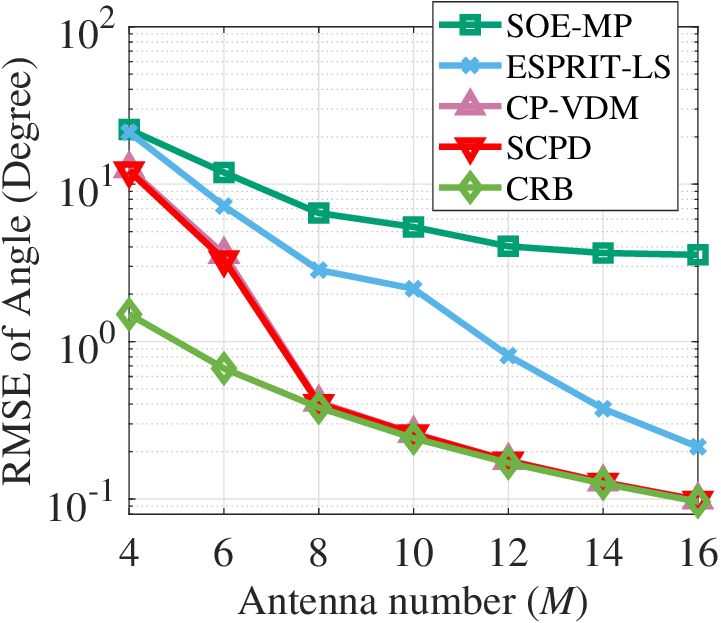}}
\caption{RMSE of parameter estimation versus antenna number $M$ in the scenario where ${\text{SNR}}=-10\ \text{dB}$ and $L=2$.}
\label{Changing_M}
\end{figure*}
\begin{figure*}[!t]
\centering
{\includegraphics[width=3.5cm]{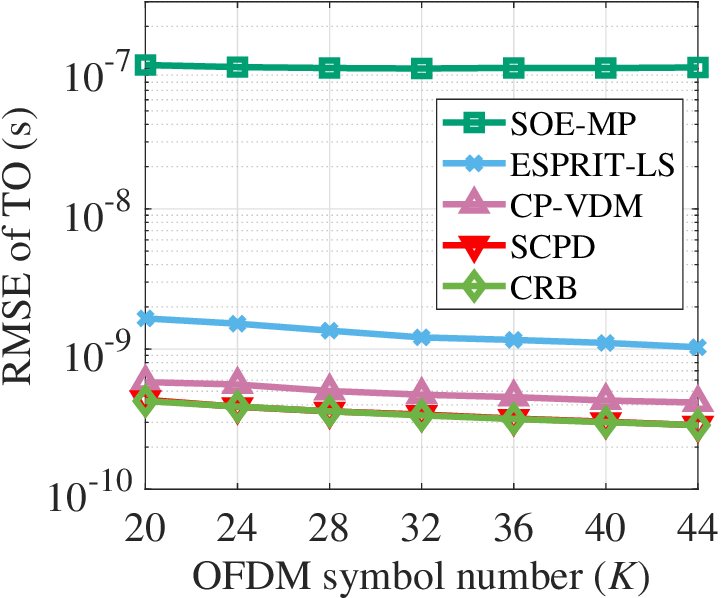}}
{\includegraphics[width=3.5cm]{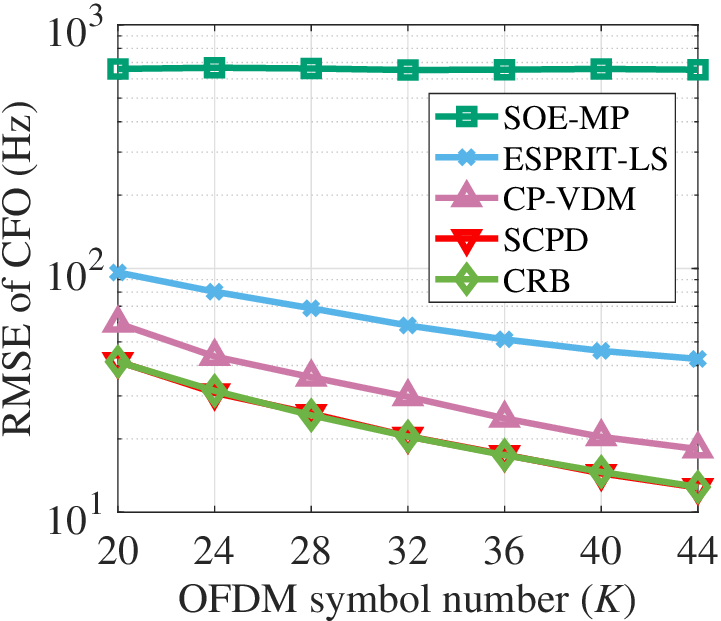}}
{\includegraphics[width=3.5cm]{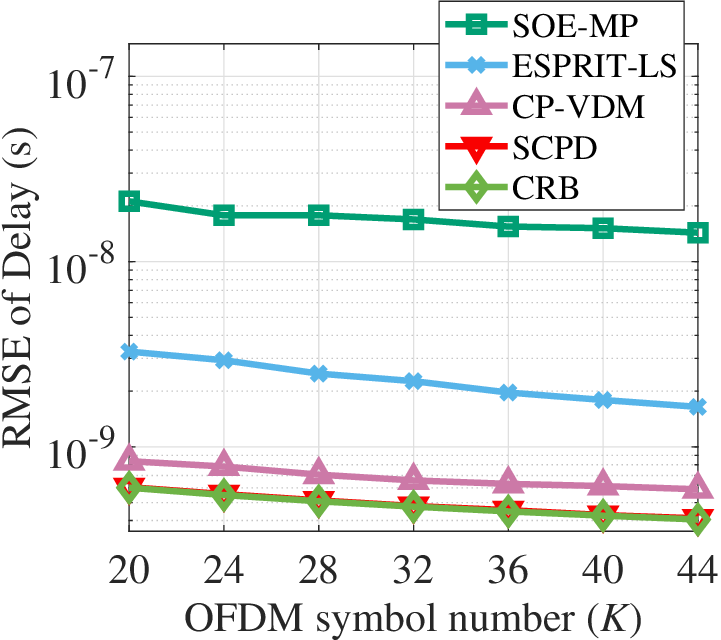}}
{\includegraphics[width=3.5cm]{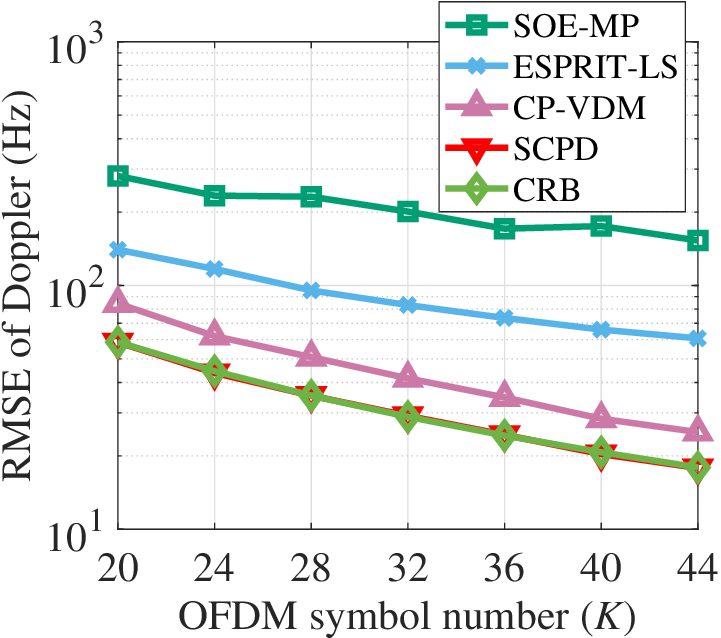}}
{\includegraphics[width=3.5cm]{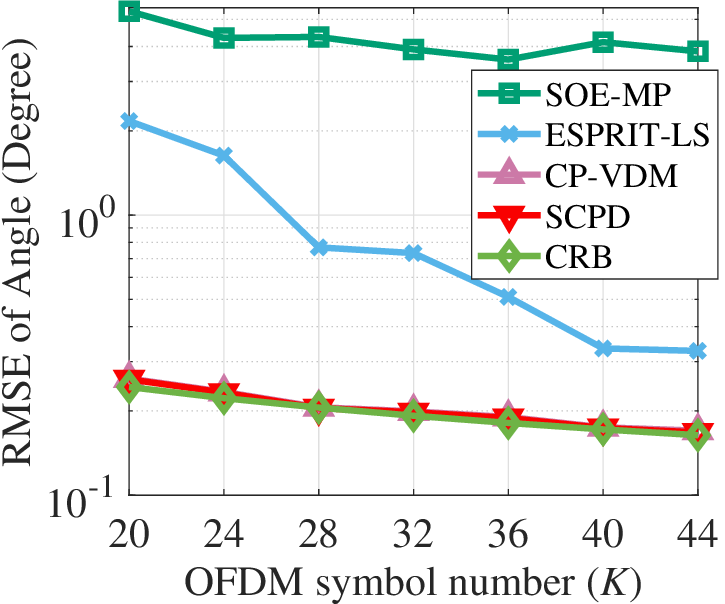}}
\caption{RMSE of parameter estimation versus OFDM symbol number $K$ in the scenario where ${\text{SNR}}=-10\ \text{dB}$ and $L=2$.}
\label{Changing_K}
\end{figure*}
\begin{figure*}[!t]
\centering
{\includegraphics[width=3.5cm]{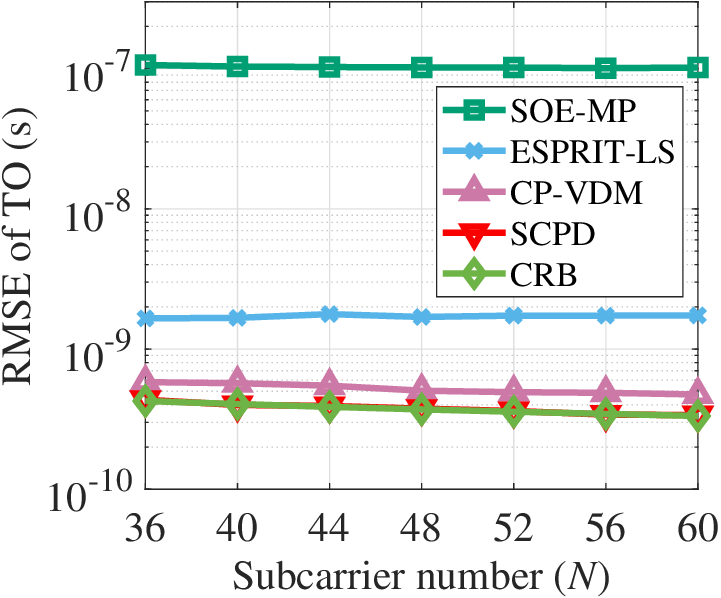}}
{\includegraphics[width=3.5cm]{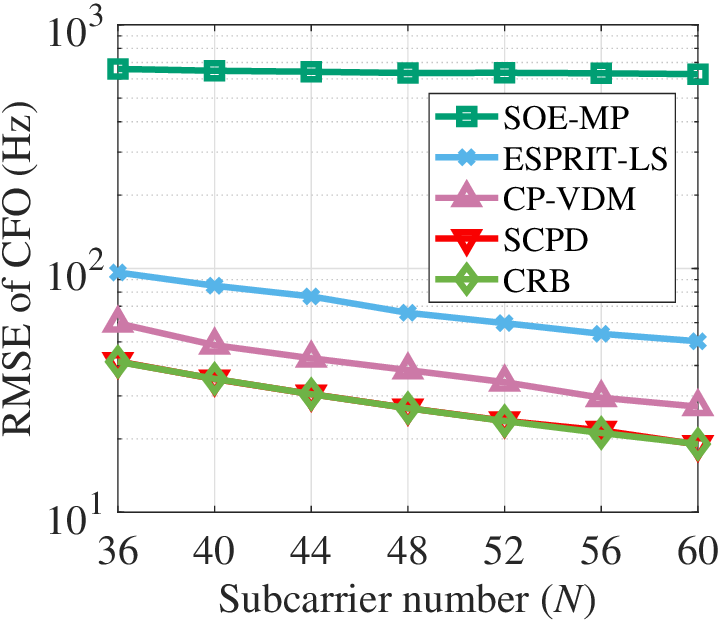}}
{\includegraphics[width=3.5cm]{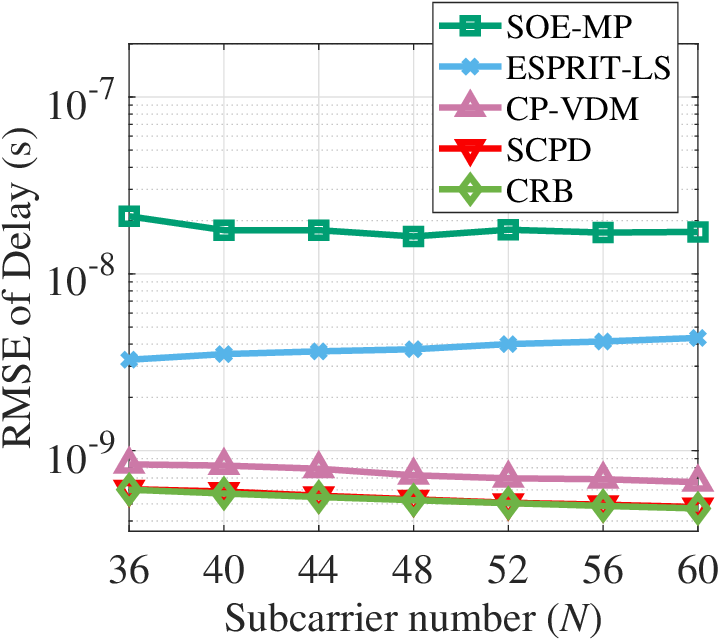}}
{\includegraphics[width=3.5cm]{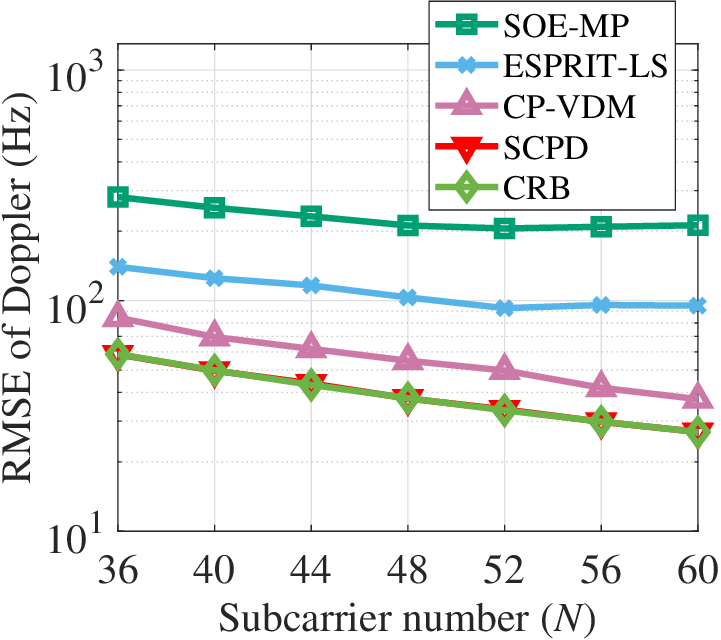}}
{\includegraphics[width=3.5cm]{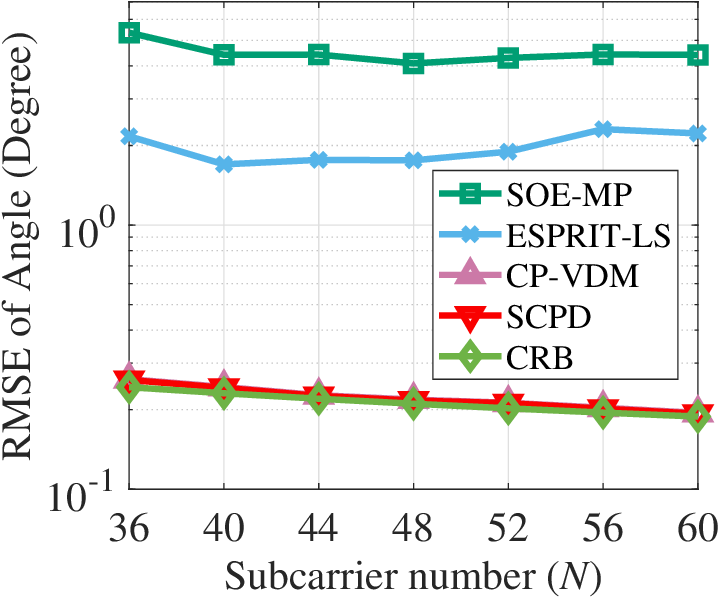}}
\caption{RMSE of parameter estimation versus subcarrier number $N$ in the scenario where ${\text{SNR}}=-10\ \text{dB}$ and $L=2$.}
\label{Changing_N}
\end{figure*}
Fig. \ref{L1_results} shows the average RMSE over all BS pairs versus SNR in the single-target scenario. The SNR at each $j$th BS pair is defined as $
\mathrm{SNR}=10\log_{10}({\|\boldsymbol{\alpha}_i\|_2^2}/{\sigma_i^2})$ for $i\in \{j_1,j_2\}$, where the vector $\boldsymbol{\alpha}_i= [\alpha_{i,1},\ldots,\alpha_{i,L}]^T$ contains the reflection coefficients of $L$ targets, as shown in \eqref{chan}, and $\sigma_i^2$ is the noise variance, as shown in \eqref{fim_eq}. Note that the RMSE results in Fig. \ref{L1_results} (and Figs. \ref{L2_results}-\ref{Changing_N}) are averaged over 3000 Monte Carlo trials, consisting of 500 trials from each of 6 BS pairs. Given that all methods involved in Figs. \ref{L1_results}-\ref{Changing_N} may fail in some Monte Carlo trials, producing RMSE outliers, the curves in Figs. \ref{L1_results}-\ref{Changing_N} are smoothed by averaging only over successful trials and excluding the failed ones. Specifically, a trial is regarded as failed if the estimated bistatic range of any target is unreliable and satisfies the first condition in \eqref{twoconditions}.
\par As shown in Fig. \ref{L1_results}, the proposed SCPD achieves the best performance in estimating both offset and target parameters. The performance gain of SCPD over SOE-MP and ESPRIT-LS is attributed to preserving the multi-dimensional structure of sensing channels. In contrast, SOE-MP and ESPRIT-LS require signal compression and unfolding, respectively, which inevitably leads to loss of structural information and performance degradation. Furthermore, SCPD approaches the CRB for ${\text{SNR}}\ge-20$ dB, demonstrating the effectiveness of using Vandermonde structural constraints for accurate tensor decomposition. Although CP-VDM also utilizes the Vandermonde structure of factor matrices in CPD, it is inferior to SCPD in terms of noise robustness. As a subspace-based method, CP-VDM is suboptimal in handling Gaussian noise and may yield biased decomposition between the signal subspace and noise subspace\cite{qian2018tensor}. To enhance noise robustness, SCPD optimizes Vandermonde factor matrices via CALS and yields maximum likelihood estimates under independent identically distributed (i.i.d.) circularly symmetric Gaussian noise. Under mild regularity conditions, maximum likelihood estimates are asymptotically unbiased and capable of reaching the CRB\cite{qian2018tensor}, as shown in Figs. \ref{L1_results}-\ref{Changing_N}.
\par Next, we examine the parameter estimation accuracy for the two-target scenario in Fig. \ref{L2_results}, following the experimental setup in Fig. \ref{L1_results}, except for the settings of $L=2$ and target locations. In Fig. \ref{L2_results}, the two targets are well-separated, with their angular interval and bistatic range interval no less than 12 degrees and 12 m, respectively. As shown in Fig. \ref{L2_results}, SOE-MP performs poorly in both offset and target parameter estimation. This is because the channel compression in SOE-MP causes the inter-path interference in scenarios where $L>1$. Specifically, the compressed vector for offset estimation in SOE-MP comprises multiple paths. Among these paths, the desired one may be indistinguishable from other interfering paths, leading to performance degeneration in offset estimation and subsequent target sensing. Unlike SOE-MP, ESPRIT-LS does not require the signal compression. It successfully separates multipath components in the sensing channel, thereby avoiding the inter-path interference. Compared with ESPRIT-LS, the proposed tensor-based framework accounts for the multi-dimensional signal structure and performs better in Figs. \ref{L1_results} and \ref{L2_results}. Furthermore, under this framework, the proposed SCPD outperforms CP-VDM due to its enhanced noise robustness, achieved by alternately and iteratively optimizing multiple factor matrices in the CPD. 
\par In addition to the estimation accuracy, we also evaluate the estimation stability of different methods using the success rate criterion. Table \ref{tab:invalid_counts} reports the success rates of parameter estimation over all 3000 Monte Carlo trials, corresponding to the RMSE results in Fig. \ref{L1_results} and \ref{L2_results}. As shown in Table \ref{tab:invalid_counts}, all methods achieve improved stability as SNR increases. The proposed SCPD attains competitive success rates across all SNR regimes in Table \ref{tab:invalid_counts}, demonstrating reasonable robustness against outliers in parameter estimation. Additionally, we test the robustness of different methods to variations in three system parameters: the antenna number ($M$), OFDM symbol number ($K$), and subcarrier number ($N$). Following the experimental setup in Fig. \ref{L2_results} at an SNR of $-10$ dB, we vary these parameters and report the average RMSE results in Figs. \ref{Changing_M}-\ref{Changing_N}. As $M$, $K$, and $N$ increase, the proposed tensor-based framework consistently benefits from the enlarged tensor dimensions and remains stable across all three parameter variations. In comparison, the performance of SOE-MP is limited by its channel compression step, while ESPRIT-LS is more sensitive to system parameter changes than CP-VDM and SCPD.
\subsection{Results on Target Tracking}
\newcommand{\trackingsubfig}[2]{%
\begin{minipage}{0.33\linewidth}
\centering
\includegraphics[width=\linewidth]{#1}\\[-0.16cm]
\makebox[\linewidth][c]{\hspace*{0.4cm}{\footnotesize #2}}
\end{minipage}}
\begin{figure}[htbp]
\centering
\begin{tabular}{@{}c@{\hspace{0.002\linewidth}}c@{\hspace{0.002\linewidth}}c@{}}
\trackingsubfig{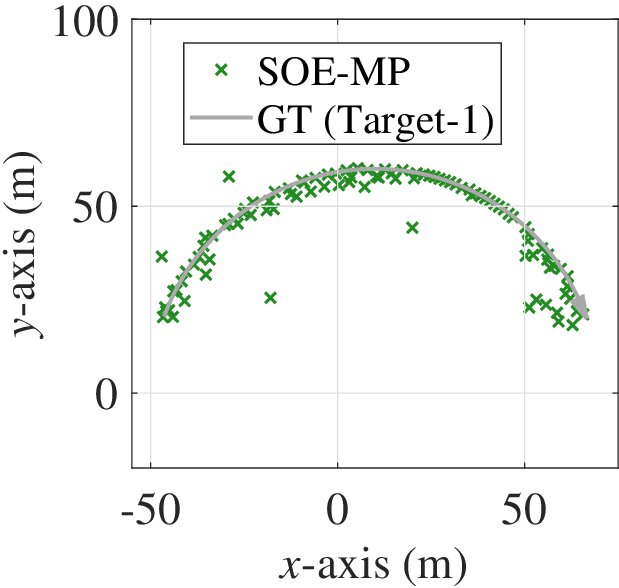}{(a)} &
\trackingsubfig{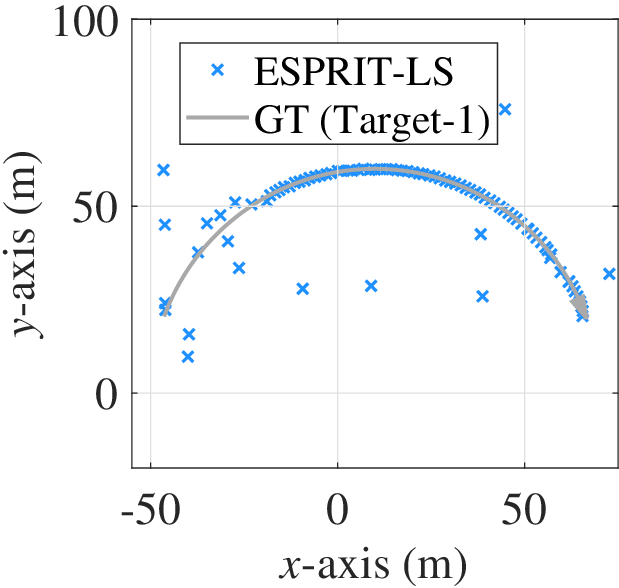}{(b)} &
\trackingsubfig{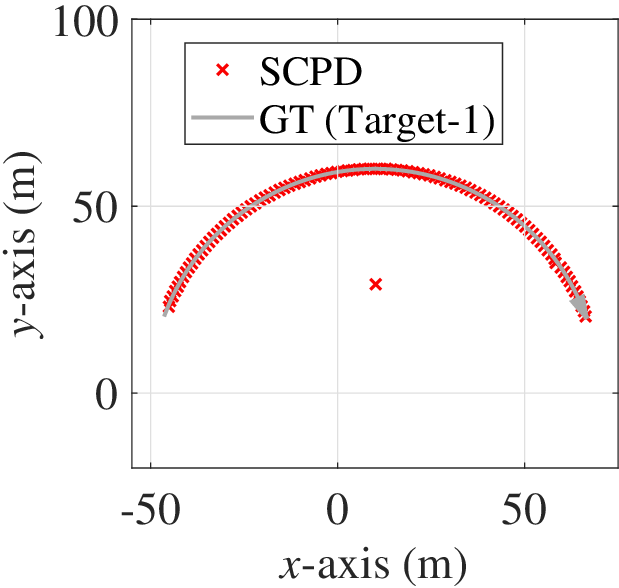}{(c)} \vspace{0.1cm} \\
\trackingsubfig{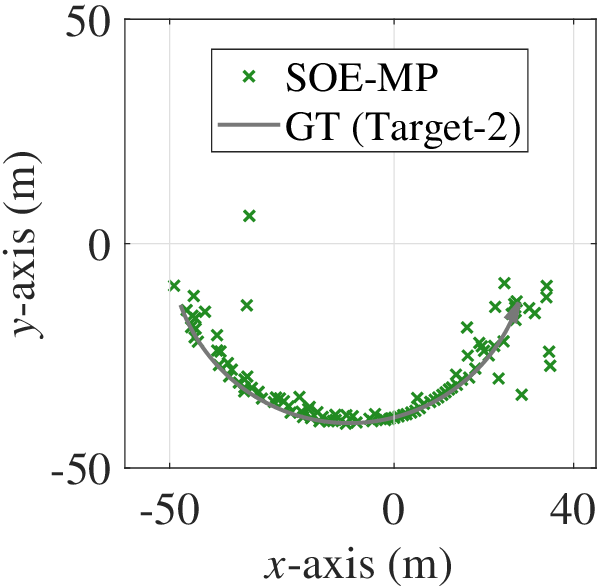}{(d)} &
\trackingsubfig{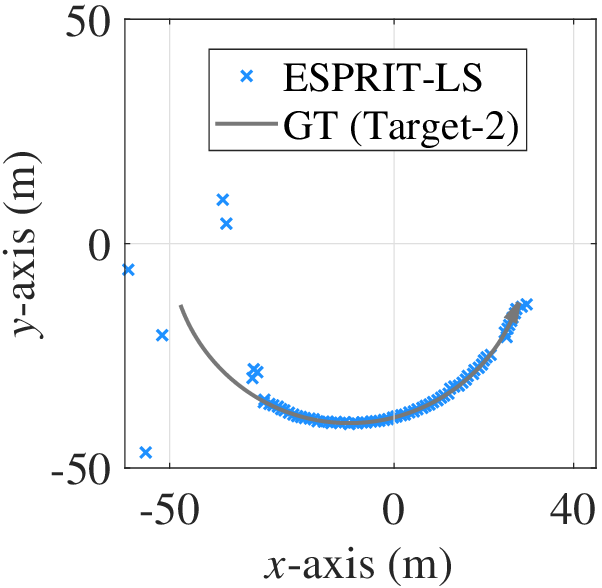}{(e)} &
\trackingsubfig{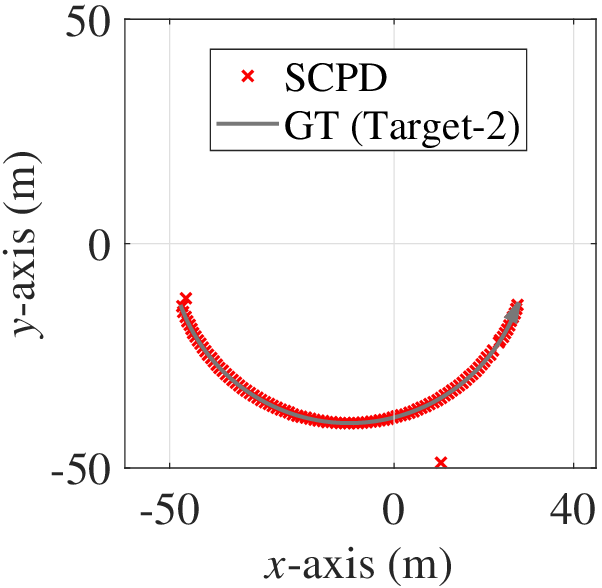}{(f)}
\end{tabular}
\caption{Trajectory tracking results in the scenario where ${\text{SNR}}=0\ \text{dB}$ and $L=2$, achieving using SOE-MP, ESPRIT-LS, and SCPD methods for target-1 in (a)-(c), and for target-2 in (d)-(f), respectively.}
\label{fig:tracking_traj_0db}
\end{figure}
\begin{figure}[htbp]
\centering
\begin{tabular}{@{}c@{\hspace{0.002\linewidth}}c@{\hspace{0.002\linewidth}}c@{}}
\trackingsubfig{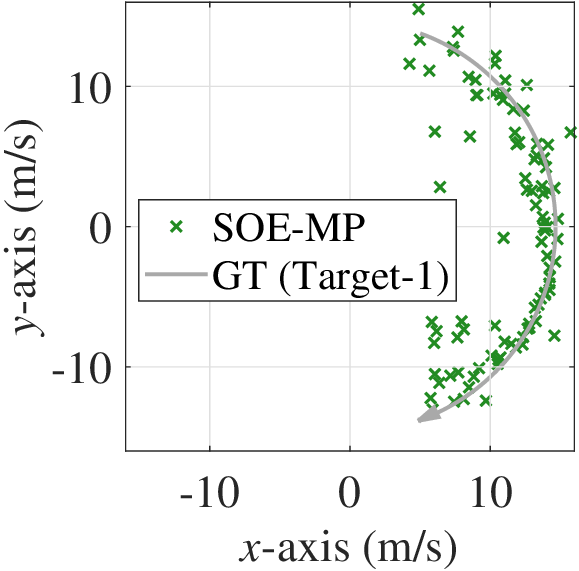}{(a)} &
\trackingsubfig{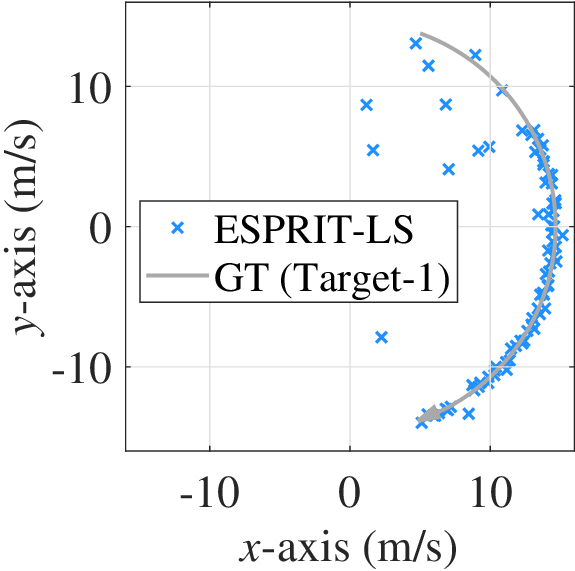}{(b)} &
\trackingsubfig{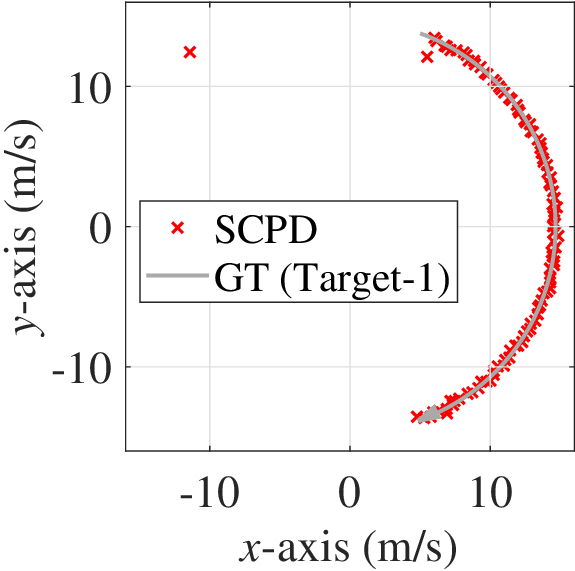}{(c)} \vspace{0.1cm} \\
\trackingsubfig{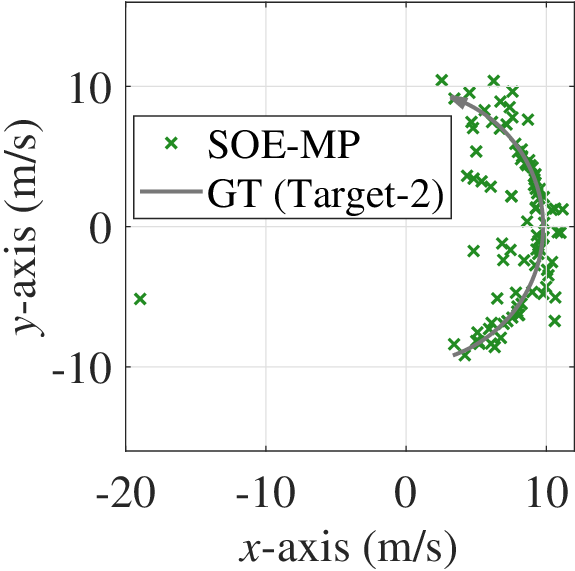}{(d)} &
\trackingsubfig{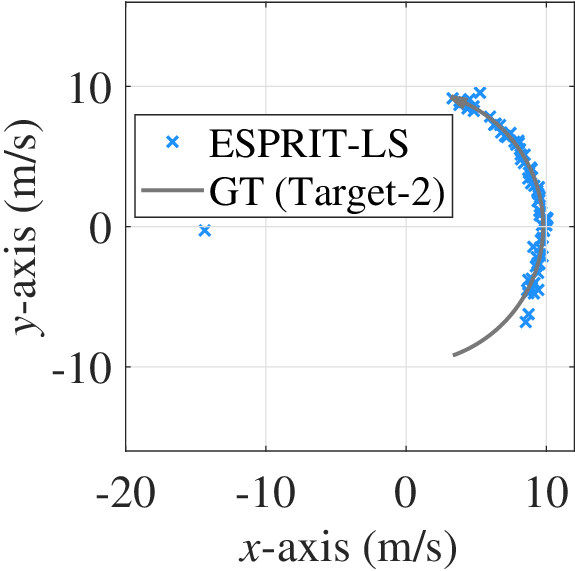}{(e)} &
\trackingsubfig{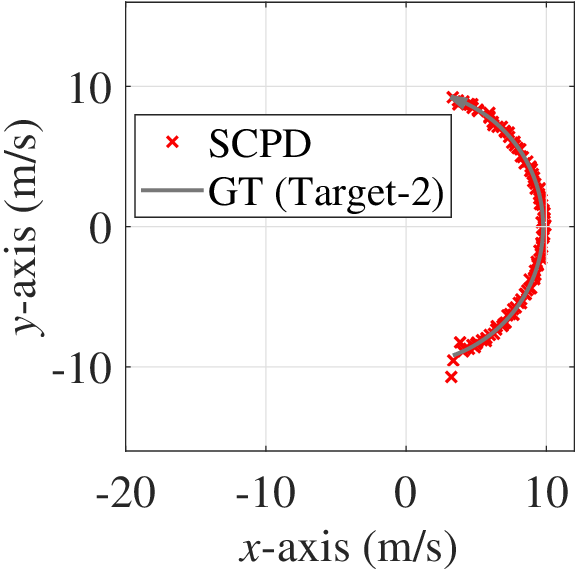}{(f)}
\end{tabular}
\caption{Velocity tracking results in the scenario where ${\text{SNR}}=0\ \text{dB}$ and $L=2$, achieving using SOE-MP, ESPRIT-LS, and SCPD methods for target-1 in (a)-(c), and for target-2 in (d)-(f), respectively.}
\label{fig:tracking_vel_0db}
\end{figure}
\begin{figure}[htbp]
\centering
\begin{tabular}{@{}c@{\hspace{0.002\linewidth}}c@{\hspace{0.002\linewidth}}c@{}}
\trackingsubfig{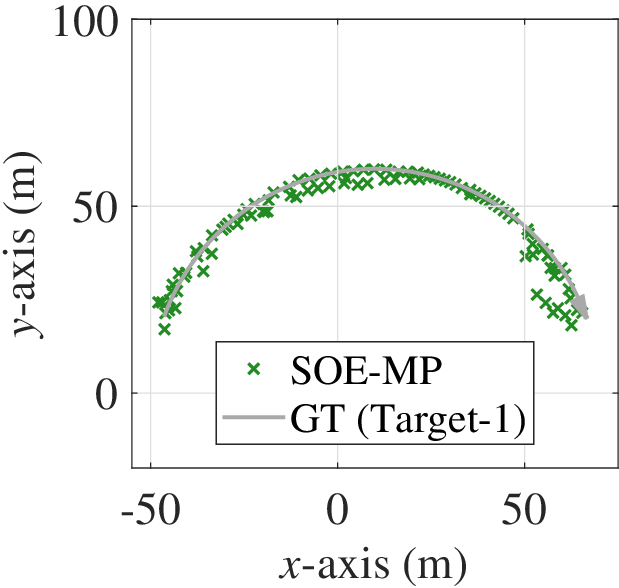}{(a)} &
\trackingsubfig{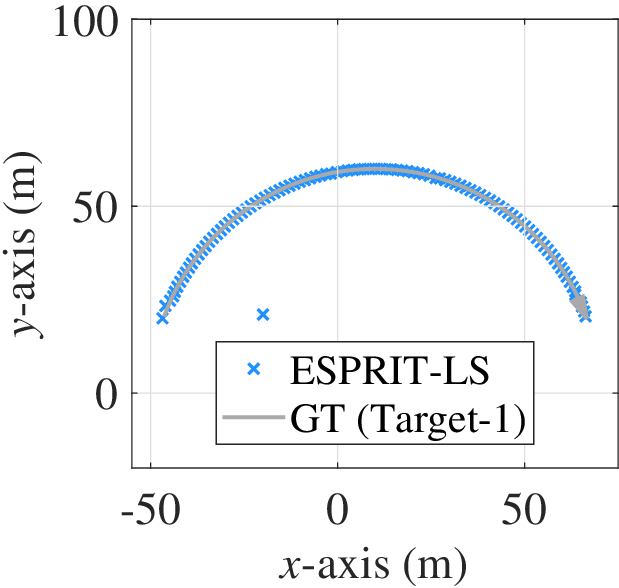}{(b)} &
\trackingsubfig{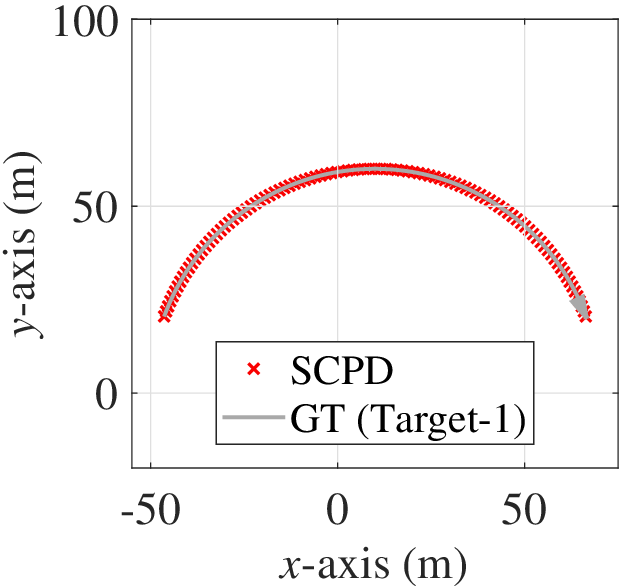}{(c)} \vspace{0.1cm} \\
\trackingsubfig{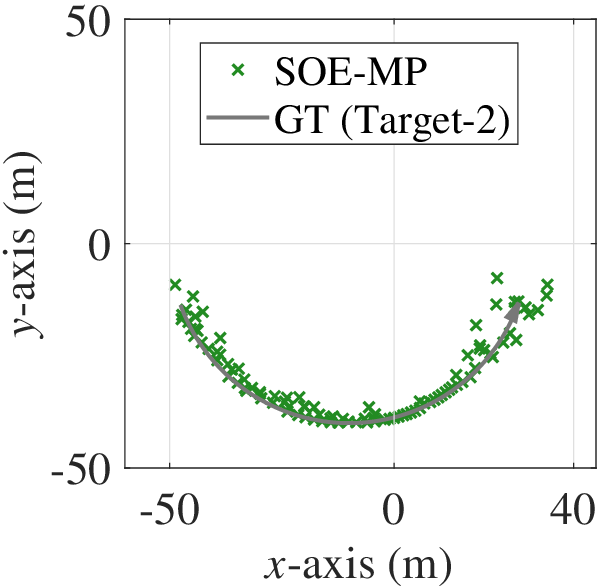}{(d)} &
\trackingsubfig{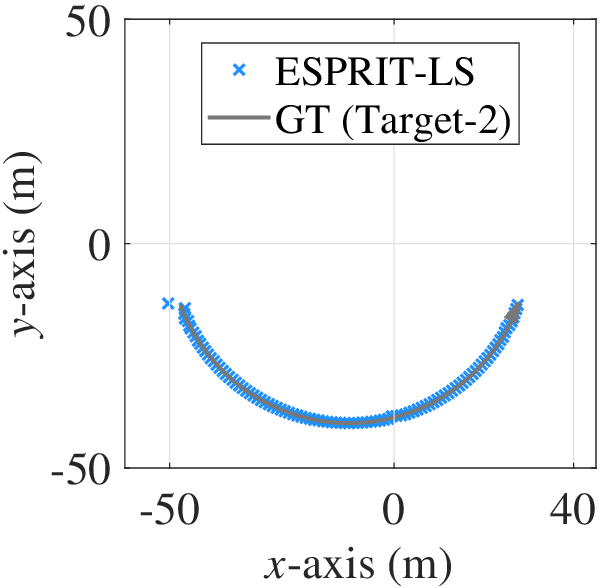}{(e)} &
\trackingsubfig{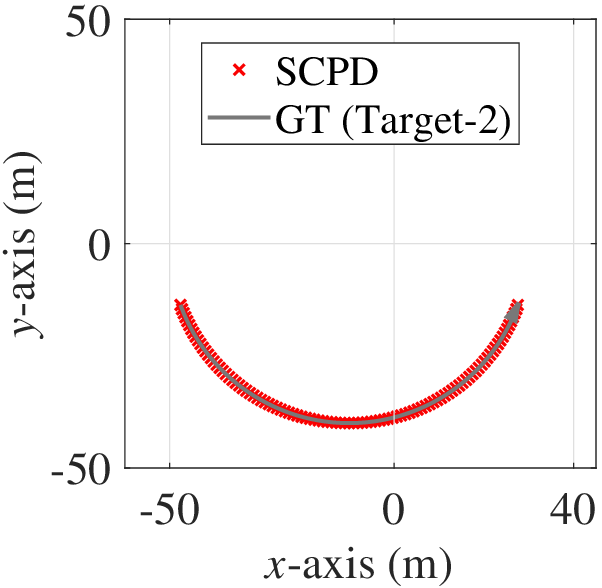}{(f)}
\end{tabular}
\caption{Trajectory tracking results in the scenario where ${\text{SNR}}=20\ \text{dB}$ and $L=2$, achieving using SOE-MP, ESPRIT-LS, and SCPD methods for target-1 in (a)-(c), and for target-2 in (d)-(f), respectively.}
\label{fig:tracking_traj_20db}
\end{figure}
\begin{figure}[htbp]
\centering
\begin{tabular}{@{}c@{\hspace{0.002\linewidth}}c@{\hspace{0.002\linewidth}}c@{}}
\trackingsubfig{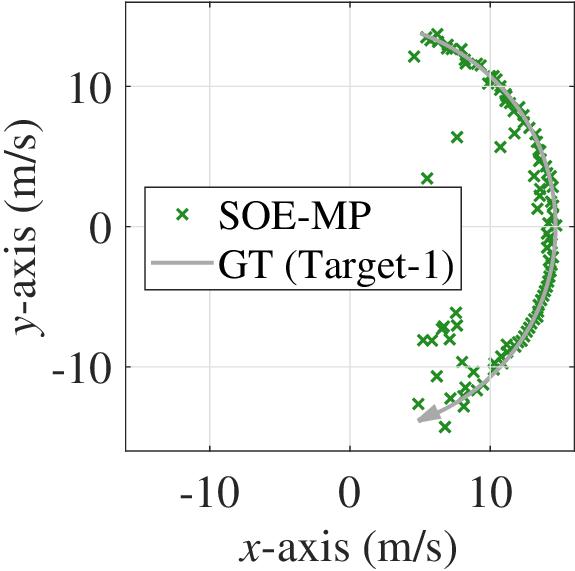}{(a)} &
\trackingsubfig{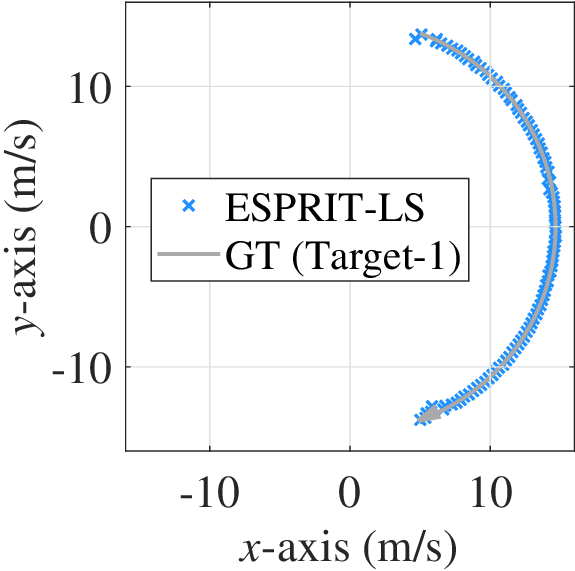}{(b)} &
\trackingsubfig{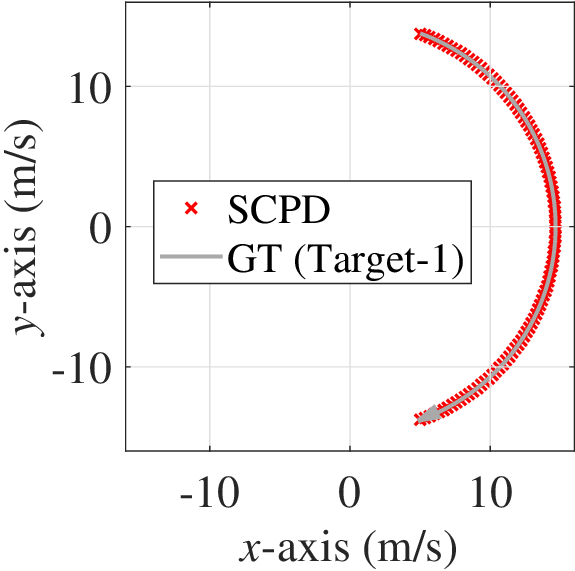}{(c)} \vspace{0.1cm} \\
\trackingsubfig{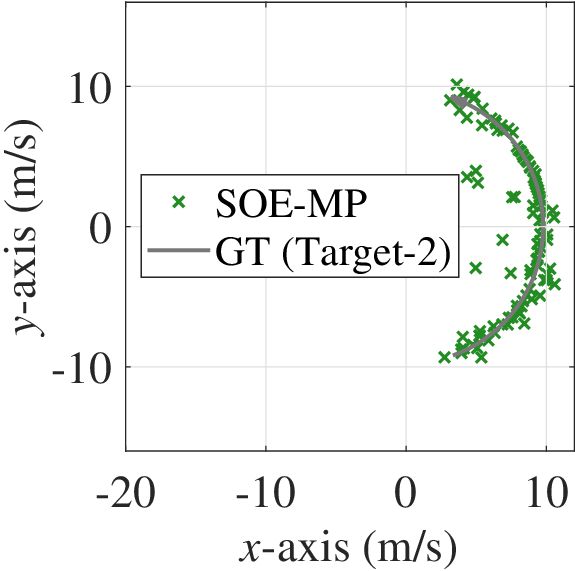}{(d)} &
\trackingsubfig{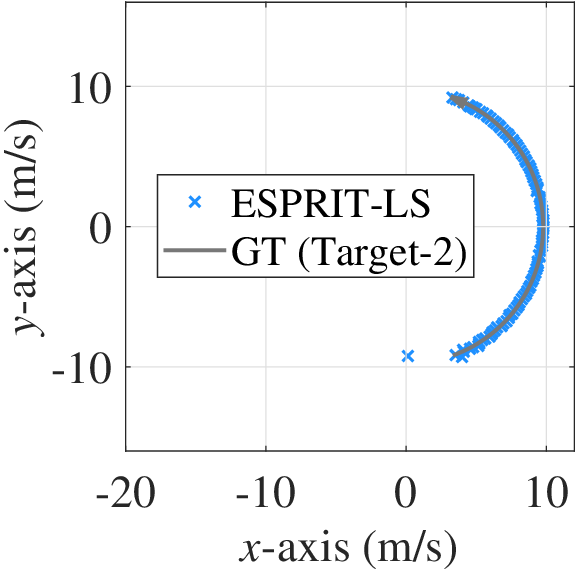}{(e)} &
\trackingsubfig{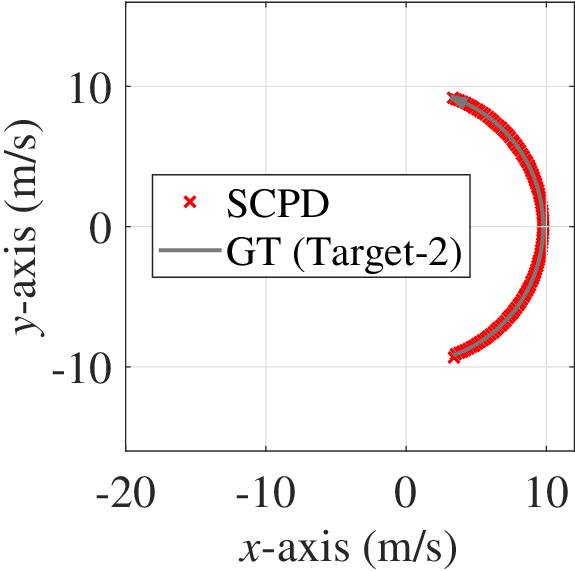}{(f)}
\end{tabular}
\caption{Velocity tracking results in the scenario where ${\text{SNR}}=20\ \text{dB}$ and $L=2$, achieving using SOE-MP, ESPRIT-LS, and SCPD methods for target-1 in (a)-(c), and for target-2 in (d)-(f), respectively.}
\label{fig:tracking_vel_20db}
\end{figure}
We evaluate the performance of SCPD on target tracking and compare it with SOE-MP and ESPRIT-LS. All three methods follow the same tracking procedure as in Algorithm \ref{alg:tracking_flow}, but use different target parameter estimates $\{\mathcal{Z}_{j,t}\}_{j=1,t=1}^{J,T}$ as algorithm inputs. The system parameters and algorithm settings follow those in Section \ref{paraest}, with a different scenario. In this scenario, two targets move continuously over $T=101$ snapshots of measurements, as described in Section \ref{beamdesign}. The trajectories and velocity variations of the two targets over time are arc-shaped, and their ground-truth (GT) values are illustrated in Figs. \ref{fig:tracking_traj_0db}-\ref{fig:tracking_vel_20db}.
\par Figs. \ref{fig:tracking_traj_0db} and \ref{fig:tracking_vel_0db} show the trajectory and velocity tracking results of different methods at an SNR of 0 dB. As shown, SCPD yields reliable tracking performance, demonstrating the effectiveness of Algorithm \ref{alg:tracking_flow}. Furthermore, SCPD outperforms SOE-MP and ESPRIT-LS in both tracking accuracy and robustness against outliers, as displayed in Figs. \ref{fig:tracking_traj_0db} and \ref{fig:tracking_vel_0db}. Apart from the visual comparison, we also compare them quantitatively in terms of RMSE. The RMSEs for target location and velocity estimation are respectively defined as ${\text{RMSE}}(\bm{u}) = \sqrt{\frac{1}{LT} \sum_{l=1,t=1}^{L,T}\|\bm{u}_{l,t}-\hat{\bm{u}}_{l,t}\|^2_2}$ and ${\text{RMSE}}(\bm{v}) = \sqrt{\frac{1}{LT} \sum_{l=1,t=1}^{L,T}\|\bm{v}_{l,t}-\hat{\bm{v}}_{l,t}\|^2_2}$. Corresponding to the visual results in Fig. \ref{fig:tracking_traj_0db}, the localization RMSEs achieved by SCPD, SOE-MP, and ESPRIT-LS are $4.58$ m, $10.77$ m, $42.53$ m, respectively. Their velocity estimation RMSEs are $16.89$ m/s, $17.80$ m/s, $93.66$ m/s, respectively, corresponding to the visual results in Fig. \ref{fig:tracking_vel_0db}. These visual and quantitative advantages of SCPD over SOE-MP and ESPRIT-LS are attributed to its more accurate target parameter estimates, which serve as inputs for tracking and adaptive beamforming.
\par Following the experimental setup in Figs. \ref{fig:tracking_traj_0db} and \ref{fig:tracking_vel_0db}, we change the SNR from 0 dB to 20 dB and display the tracking results in Figs. \ref{fig:tracking_traj_20db} and \ref{fig:tracking_vel_20db}. As shown, SCPD yields reasonable tracking results that are close to their GT values. It achieves localization and velocity estimation RMSEs of $0.008$ m and $0.018$ m/s, respectively. In comparison, SOE-MP results in larger RMSEs of $3.62$ m for localization and $1.84$ m/s for velocity estimation. ESPRIT-LS attains a localization RMSE of $1.84$ m, but a high RMSE for velocity estimation of $13.38$ m/s, due to its unsatisfactory robustness against outliers. These RMSE results show that, when the target parameter estimates become more accurate, Algorithm \ref{alg:tracking_flow} achieves more reliable tracking of target trajectories and velocities.
\section{Conclusion}
This paper proposes an integrated framework for joint time-frequency synchronization and multi-target sensing in networked ISAC. In this framework, SCPD is employed to estimate pairwise offset and target parameters. By decomposing the multi-dimensional sensing channel into Vandermonde factor matrices, SCPD isolates multipath components to avoid inter-path interference. The optimization problem of SCPD is effectively solved via the CALS approach, and its uniqueness conditions as well as the CRB for parameter estimation are established. Building upon the pairwise parameter estimates, a target tracking algorithm is further proposed, which associates and fuses estimates from multiple BS pairs to recover both 2D trajectories and 2D velocities of moving targets. To enhance tracking performance, an adaptive beamforming scheme is developed based on the prior trajectory estimates to illuminate the predicted target directions. Simulation results validate the effectiveness of the proposed synchronization and sensing framework, demonstrating that SCPD achieves superior accuracy and outlier robustness in parameter estimation and target tracking compared to traditional approaches.
\begin{appendices}
\section{SOE-MP Extension for Target Sensing}
\label{app:soempm_sic}
This appendix extends SOE-MP to estimate target parameters in the MIMO scenario using an iterative scheme. This scheme initializes an angle-delay-Doppler spectrum $\pmb{\mathcal{R}}^1_i\triangleq \pmb{\mathcal{E}}_i$, as shown in \eqref{eq:soempm_transform}, containing $L$ spectral components corresponding to $L$ targets. Then, in each $l$th iteration, SOE-MP estimates the $l$th target's parameters from the current spectrum $\pmb{\mathcal{R}}^l_i$, and reconstructs the spectral component relative to the $l$th target. This reconstructed component is then subtracted from $\pmb{\mathcal{R}}^l_i$, updating it to $\pmb{\mathcal{R}}^{l+1}_i$ for the $(l+1)$th iteration.
\par Specifically, to estimate the $l$th target's parameters from $\pmb{\mathcal{R}}^l_i$, we select an angle-delay-Doppler bin in $\pmb{\mathcal{R}}^l_i$ by maximizing the normalized energy:
\begin{equation}
(m_{i,l},n_{i,l},k_{i,l})
=
{\arg\max\limits_{(m,n,k)}}\ h\big(m,n,k \,\big|\, \pmb{\mathcal{R}}^l_i\big),
\label{eq:soempm_sic_triplet}
\end{equation}
where the function $h$ is defined as
\begin{equation}
\begin{aligned}
&  \ \ \ \  h\big(m,n,k \,\big|\, \pmb{\mathcal{R}}^l_i\big) \\
& \triangleq  \frac{\big\|[\pmb{\mathcal{R}}^l_i]_{m,:,k}\big\|_2^2}{\max\limits_{(a,c)} \big\|[\pmb{\mathcal{R}}^l_i]_{a,:,c}\big\|_2^2}
+ \frac{\big\|[\pmb{\mathcal{R}}^l_i]_{m,n,:}\big\|_2^2}{\max\limits_{(a,b)} \big\|[\pmb{\mathcal{R}}^l_i]_{a,b,:}\big\|_2^2}
+ \frac{\big\|[\pmb{\mathcal{R}}^l_i]_{:,n,k}\big\|_2^2}
{\max\limits_{(b,c)}  \big\|[\pmb{\mathcal{R}}^l_i]_{:,b,c}\big\|_2^2}.
\end{aligned}
\end{equation}
Under this selected bin, we estimate the parameters of the $l$th target in a manner similar to \eqref{eq:soemp_offsets} as
\begin{equation}
\!\left\{\begin{aligned}
& [\hat{\bm{\tau}}_i]_l =
- \frac{\angle\big( \bm{F}_{N}\cdot
[\pmb{\mathcal{R}}^l_i]_{m_{i,l},:,k_{i,l}} \big)}{2\pi\Delta f}, \ \forall i \in \{j_1,j_2\},\\
& [\hat{\bm{\nu}}_i]_l  =
\frac{\angle\big(\bm{F}_{K}^{-1}\cdot
[\pmb{\mathcal{R}}^l_i]_{m_{i,l},n_{i,l},:}\big)}{2\pi T_{\mathrm{sym}}}, \ \forall i \in \{j_1,j_2\},\\
& \hat{\theta}_{i,l}=
\arcsin\Bigg(\frac{\lambda  \cdot \angle\big(\bm{F}_{M}^{-1}\cdot
[\pmb{\mathcal{R}}^l_i]_{:,n_{i,l},k_{i,l}}\big)}{2\pi d}\Bigg), \ \forall i \in \{j_1,j_2\},
\end{aligned}\right.
\label{eq:soempm_sic_composite}
\end{equation}
where $\hat{\bm{\tau}}_i\in \mathbb{R}^L$ and $\hat{\bm{\nu}}_i\in \mathbb{R}^L$, with $[\hat{\bm{\tau}}_i]_l$ and $[\hat{\bm{\nu}}_i]_l$ being the estimates of the true parameters $(\tau_{j,l}+\Delta \tau_i)$ and $(\nu_{j,l}+\Delta \nu_i)$, respectively, for all $l$ and $i \in \{j_1,j_2\}$. After calculating \eqref{eq:soempm_sic_composite}, we reconstruct the spectral component $\pmb{\mathcal{A}}^l_i\in\mathbb{C}^{M\times N\times K}$ relative to the $l$th target as 
\begin{equation}
[\pmb{\mathcal{A}}^l_i]_{m,n,k}
=
[\bm{F}_{M}\cdot \bm{a}_{\mathrm{r}}(\hat{\theta}_{i,l})]_{m} \cdot 
\big[\bm{F}_{N}^{-1}\cdot [\hat{\bm{B}}_i]_{:,l} \big]_{n} \cdot 
\big[\bm{F}_{K}\cdot [\hat{\bm{C}}_i]_{:,l}\big]_{k},
\label{eq:soempm_sic_atom}
\end{equation}
where $\bm{a}_{\mathrm{r}}(\hat{\theta}_{i,l})$ is defined similarly to $\bm{a}_{\mathrm{r}}(\theta_{i,l})$ in \eqref{mknewnew}. $[\hat{\bm{B}}_i]_{:,l}$ is defined similarly to $[\bm{B}_i]_{:,l}$ in \eqref{SCPD_offset_recovery} by replacing the true parameter $(\tau_{j,l}+\Delta \tau_i)$ with its estimate $[\hat{\bm{\tau}}_i]_l$. $[\hat{\bm{C}}_i]_{:,l}$ is defined similarly to $[\bm{C}_i]_{:,l}$ in \eqref{SCPD_offset_recovery} by replacing $(\nu_{j,l}+\Delta \nu_i)$ with $[\hat{\bm{\nu}}_i]_l$. Subsequently, we cancel $\pmb{\mathcal{A}}^l_i$ from
$\pmb{\mathcal{R}}^l_i$ to yield $\pmb{\mathcal{R}}^{l+1}_i$ as \begin{equation}
\pmb{\mathcal{R}}_i^{l+1}
=\pmb{\mathcal{R}}^l_i
-\alpha^l_i\cdot \pmb{\mathcal{A}}^l_i, \ \forall i \in \{j_1,j_2\},
\label{eq:soempm_sic_update}
\end{equation}
where $\alpha^l_i=
{\left\langle \pmb{\mathcal{A}}^l_i,
\pmb{\mathcal{R}}^l_i\right\rangle}
/{\|\pmb{\mathcal{A}}^l_i\|_{F}^{2}}$ denotes the estimated amplitude of the spectral component $\pmb{\mathcal{A}}^l_i$ via the LS criterion.
\par After repeating \eqref{eq:soempm_sic_triplet}-\eqref{eq:soempm_sic_update} for all $l\in \{1,\ldots,L\}$ and $i \in \{j_1,j_2\}$, the delay and Doppler shift parameters of $L$ targets are estimated based on the bistatic reciprocity:
\begin{equation}
\!\hat{\tau}_{j,l}\!=\! \frac{\!\angle\big( [\hat{\bm{B}}_{j_1}]_{:,l}
\!\circledast\! [\hat{\bm{B}}_{j_2}]_{:,[\bm{o}]_l}\big)\!}{-4\pi\Delta f},\, 
\hat{\nu}_{j,l}\!=\! \frac{\! \angle\big([\hat{\bm{C}}_{j_1}]_{:,l}
\!\circledast\! [\hat{\bm{C}}_{j_2}]_{:,[\bm{o}]_l}\big)\!}{4\pi T_{\mathrm{sym}}},\forall l,
\label{eq:soempm_final_target_params}
\end{equation}
where the permutation vector $\bm{o}\in \mathbb{N}^L$ in \eqref{eq:soempm_final_target_params} is obtained in a manner similar to \eqref{eq:asso} as 
\begin{equation}
\begin{aligned}
& \min_{\bm{o}\in\mathbb{O}}
\Big\{\mathcal{V}\big(\hat{\bm{\tau}}_{j_1} -\hat{\bm{\tau}}_{j_2}(\bm{o}) +2 \Delta\hat{\tau}_{j_2}\cdot \bm{1}_L\big)\cdot (\Delta f)^2\\
& \ \ \ \ \ \ \ +\mathcal{V}\big(\hat{\bm{\nu}}_{j_1}\! - \! \hat{\bm{\nu}}_{j_2}(\bm{o})+2 \Delta\hat{\nu}_{j_2}\cdot \bm{1}_L\big) \cdot T_{\mathrm{sym}}^2\Big\}.
\end{aligned}
\label{eq:soempm_candidate_pairing}
\end{equation}
\section{ESPRIT-LS Estimator}
\label{sec:esprit_ls}
We derive the ESPRIT-LS estimator given the measurement tensors $(\pmb{\mathcal{Y}}_i \oslash \pmb{\mathcal{S}}_i)\in \mathbb{C}^{M\times N K}$ for $i\in \{j_1,j_2\}$ at the $j$th BS pair. ESPRIT-LS unfolds the measurement tensor along mode-2 into the matrix $(\bm{Y}_i^{(2)} \oslash \bm{S}_i^{(2)})\in \mathbb{C}^{N\times (MK)}$, and then computes its rank-$L$ truncated SVD. The resulting left singular matrix is denoted as $\bm{U}_i\in \mathbb{C}^{N\times L}$. Notably, in the noiseless case, there exists a nonsingular matrix $\bm{O}_i\in \mathbb{C}^{L\times L}$ such that
$\bm{U}_i\bm{O}_i=\bm{B}_i\in\mathbb{C}^{N\times L}$, where $\bm{B}_i$ is the delay steering matrix defined in \eqref{SCPD_offset_recovery}. The shift invariance property of $\bm{B}_i$ leads to the following EVD:
\begin{equation}
{\underline{\bm{U}}}_i^{\dagger}\cdot \overline{\bm{U}}_i=
\bm{O}_i\bm{\Lambda}_i\bm{O}_i^{-1}, \ \forall i \in \{j_1,j_2\},
\end{equation}
where $\bm{\Lambda}_i=\operatorname{diag}\big(e^{-j2\pi\Delta f(\tau_1+\Delta\tau_i)},
\ldots, e^{-j2\pi\Delta f(\tau_{j,l}+\Delta\tau_i)}\big)$. Accordingly, the sum of the delay and TO, i.e., $\{\tau_{j,l}+\Delta \tau_i\}_{l=1}^L$, can be estimated from $\bm{\Lambda}_i$ as
\begin{equation}
[\hat{\bm{\tau}}_i]_l = - \frac{\angle\big([\bm{\Lambda}_i]_{l,l}\big)}{2\pi\Delta f}, \  \forall l, \forall i \in \{j_1,j_2\},
\label{delaymatrix}
\end{equation}
where $\hat{\bm{\tau}}_i\in \mathbb{R}^L$ is defined as in \eqref{eq:soempm_sic_composite}. Based on \eqref{delaymatrix}, the matrix $\hat{\bm{B}}_i$ can be constructed similarly to $\bm{B}_i$ in \eqref{SCPD_offset_recovery} by replacing the true parameter $(\tau_{j,l}+\Delta \tau_i)$ with its estimate $[\hat{\bm{\tau}}_i]_l$, $\forall l$.
\par According to the signal model $\bm{Y}_i^{(2)} \oslash \bm{S}_i^{(2)}=(\bm{C}_i\odot \bm{A}_i)\bm{B}_i^T$ in the noiseless case, which follows from \eqref{eq1}, the matrix $(\bm{C}_i\odot \bm{A}_i)\in \mathbb{C}^{(MK)\times L}$ can be estimated via LS as
\begin{equation}
\hat{\bm{C}}_i\odot \hat{\bm{A}}_i=(\hat{\bm{B}}_i^T)^{\dagger}(\bm{Y}^{(2)}_i\oslash \bm{S}_i^{(2)}), \ \forall i \in \{j_1,j_2\}.
\label{eq:esprit_ls_sep}
\end{equation}
Then, the $l$th column of $(\hat{\bm{C}}_i\odot \hat{\bm{A}}_i)$ is reshaped into an $M\times K$ matrix, and the rank-1 truncated SVD of this matrix yields a left singular vector $\bm{p}_{i,l}\in \mathbb{C}^M$ and a right singular vector $\bm{q}_{i,l}\in \mathbb{C}^K$. Due to the shift invariance property of 
$\bm{A}_i$ and $\bm{C}_i$, as shown in \eqref{SCPD_offset_recovery}, the angle parameters $\{\theta_{i,l}\}_{l=1}^L$ and the sum of the Doppler shift and CFO, i.e., $\{\nu_{j,l}+\Delta \nu_i\}_{l=1}^L$, can be estimated from $\bm{p}_{i,l}$ and $\bm{q}_{i,l}$, respectively, as
\begin{equation}
\!\left\{\begin{aligned}
& \hat{\theta}_{i,l}= \arcsin\!\Bigg(\frac{\lambda \! \cdot\! \angle\Big(
{\underline{\bm{p}}_{i,l}^H\cdot \overline{\bm{p}}_{i,l}}
\big/{\big\|\underline{\bm{p}}_{i,l}\big\|_2^2}
\Big)}{2\pi d}\Bigg), \  \forall l, \forall i \in \{j_1,j_2\},\\
& [\hat{\bm{\nu}}_i]_l= \frac{\angle\Big(
{\underline{\bm{q}}_{i,l}^H\cdot \overline{\bm{q}}_{i,l}}
\big/{\big\|\underline{\bm{q}}_{i,l}\big\|_2^2}
\Big)}{2\pi T_{\mathrm{sym}}}, \  \forall l, \forall i \in \{j_1,j_2\},
\end{aligned}\right.
\label{twos}
\end{equation}
where $\hat{\bm{\nu}}_i\in \mathbb{R}^L$ is defined as in \eqref{eq:soempm_sic_composite}.
\par Based on \eqref{delaymatrix} and \eqref{twos}, we employ the bistatic reciprocity to estimate TO and CFO as
\begin{equation}
{\Delta \hat{\tau}_{j_2}}= \frac{\mathcal{M}\big(\hat{\bm{\tau}}_{j_2}-\hat{\bm{\tau}}_{j_1}\big)}{2}, \quad {\Delta \hat{\nu}_{j_2}}= \frac{\mathcal{M}\big(\hat{\bm{\nu}}_{j_2}-\hat{\bm{\nu}}_{j_1}\big)}{2},
\label{eq:esprit_offsets}
\end{equation}
and to estimate delay and Doppler
shift parameters as
\begin{equation}
\hat{\tau}_{j,l}=
\frac{[\hat{\bm{\tau}}_{j_1}]_l+[\hat{\bm{\tau}}_{j_2}(\bm{o})]_l}{2}, \quad
\hat{\nu}_{j,l}=
\frac{[\hat{\bm{\nu}}_{j_1}]_l+[\hat{\bm{\nu}}_{j_2}(\bm{o})]_l}{2}, \ \forall l,
\label{eq:esprit_absolute}
\end{equation}
where the permutation vector $\bm{o}\in \mathbb{N}^L$ is obtained from \eqref{eq:soempm_candidate_pairing}.
\end{appendices}
\bibliographystyle{IEEEtran}
\bibliography{Bibliography}
\end{document}